\documentclass[fleqn,usenatbib]{mnras}

\DeclareRobustCommand{\VAN}[3]{#2}
\let\VANthebibliography\thebibliography
\def\thebibliography{\DeclareRobustCommand{\VAN}[3]{##3}\VANthebibliography}

\usepackage{enumitem}
\setlist[enumerate]{labelwidth=*, leftmargin=2em}
\usepackage[T1]{fontenc}
\usepackage{amsmath}
\usepackage{amssymb}
\usepackage{graphicx}
\usepackage{booktabs}
\usepackage{lmodern}
\usepackage{multicol}  
\usepackage[]{xcolor}
\usepackage{threeparttable}
\usepackage{float}
\usepackage{tabularx}
\newcolumntype{Y}{>{\centering\arraybackslash}X}

\newcommand{\RNum}[1]{\uppercase\expandafter{\romannumeral #1\relax}}

\title[Galaxy cluster thermodynamics at $z=1.7$]{Galaxy cluster thermodynamics at $z=1.7$: Resolved tSZ and X-ray analysis of CL~J0459.6-4946}

\author[van Marrewijk et al.]{J.~van Marrewijk,$^{1}$\thanks{E-mail: marrewijk@strw.leidenuniv.nl}
L.~Di~Mascolo,$^{2}$
S.~Ueda,$^{3,4}$
T.~Mroczkowski,$^{5,6}$
T.~Akahori,$^{7,8}$
\newauthor
E.~S.~Battistelli,$^{9,10,11}$
M.~J.~Devlin,$^{12}$
A.~Flores,$^{13}$
A.~S.~Gill,$^{14}$
F.~A.~G\'omez,$^{15}$
M.~Hilton,$^{16,17}$
\newauthor
J.~P.~Hughes,$^{13}$
T.~Kitayama,$^{18}$
Y.-T.~Lin,$^{19}$
K.~Moodley,$^{20,21}$
N.~Okabe,$^{22,23,24}$
\newauthor
M.~Pannella,$^{25,26}$
B.~Ried~Guachalla,$^{27,28,29}$
H.~R\"ottgering,$^{1}$
J.~S.~Sanders,$^{30}$
A.~Saro,$^{25,26,31,32,33}$
\newauthor
C.~Sif\'on,$^{34}$
V.~Strazzullo,$^{25,26}$
M.~Takizawa,$^{35}$
E.~J.~Wollack,$^{36}$
and X.~Zhang$^{30}$ \\
\emph{Affiliations are listed at the end of the paper}
}

\newcommand{\affiliations}{%
  \section*{Affiliations}%
\noindent {\it
$^{1}$Leiden Observatory, Leiden University, P.O. Box 9513, 2300 RA Leiden, The Netherlands\\
$^{2}$Kapteyn Astronomical Institute, University of Groningen, Landleven 12, 9747 AD, Groningen, The Netherlands \\
$^{3}$Faculty of Mathematics and Physics, Institute of Science and Engineering, Kanazawa University, Kakuma, Kanazawa, Ishikawa, 920-1192, Japan \\
$^{4}$Advanced Research Center for Space Science and Technology (ARC-SAT), Kanazawa University, Kakuma, Kanazawa, Ishikawa, 920-1192, Japan \\
$^{5}$Institute of Space Sciences (ICE-CSIC), Carrer de Can Magrans, s/n, 08193 Cerdanyola del Vall\`es, Barcelona, Spain\\
$^{6}$Institut d'Estudis Espacials de Catalunya (IEEC), E-08860 Castelldefels, Barcelona, Spain\\
$^{7}$Mizusawa VLBI Observatory, National Astronomical Observatory of Japan, 2-21-1 Mitaka, Tokyo 181-8588, Japan \\
$^{8}$SOKENDAI, Shonan Village, Hayama-machi, Miura-gun, Kanagawa 240-0193, Japan \\
$^{9}$Sapienza – University of Rome - Physics department, Piazzale Aldo Moro 5 00185 Rome, Italy \\
$^{10}$INFN – Rome Division, Piazzale Aldo Moro 5 00185 Rome, Italy \\
$^{11}$INAF – IAPS,  Via del Fosso del Cavaliere 100, 00133 Rome, Italy \\
$^{12}$Department of Physics and Astronomy, University of Pennsylvania, 209 South 33rd Street, Philadelphia, PA, USA 19104 \\
$^{13}$Department of Physics and Astronomy, Rutgers, The State University of New Jersey, Piscataway, NJ USA 08854-8019 \\
$^{14}$National Research Council Canada Herzberg Astronomy and Astrophysics Research Centre, Victoria, BC, V9E 2E7 Canada \\
$^{15}$Departamento de Astronom\'ia, Universidad de La Serena, Av. Ra\'ul Bitr\'an 1305, La Serena, Chile \\
$^{16}$Wits Centre for Astrophysics, School of Physics, University of the Witwatersrand, Private Bag 3, 2050, Johannesburg, South Africa \\
$^{17}$School of Mathematics, Statistics, and Computer Science, University of KwaZulu-Natal, Westville Campus, Durban 4041, South Africa \\
$^{18}$Department of Physics, Toho University, 2-2-1 Miyama, Funabashi, Chiba 274-8510, Japan\\
$^{19}$Academia Sinica Institute of Astronomy and Astrophysics (ASIAA), No. 1, Section 4, Roosevelt Road, Taipei 10617, Taiwan \\
$^{20}$A. Astrophysics Research Centre, University of KwaZulu-Natal, Westville Campus, Durban 4041, South Africa \\
$^{21}$B. School of Agriculture and Science, University of KwaZulu-Natal, Westville Campus, Durban 4041, South Africa \\
$^{22}$Physics Program, Graduate School of Advanced Science and Engineering, Hiroshima University, 1-3-1 Kagamiyama, Higashi-Hiroshima, Hiroshima 739-8526, Japan \\
$^{23}$Hiroshima Astrophysical Science Center, Hiroshima University, 1-3-1 Kagamiyama, Higashi-Hiroshima, Hiroshima 739-8526, Japan \\
$^{24}$Core Research for Energetic Universe, Hiroshima University, 1-3-1 Kagamiyama, Higashi-Hiroshima, Hiroshima 739-8526, Japan \\
$^{25}$INAF -- Observatory of Trieste, Via G. Tiepolo 11, I-34143 Trieste, Italy \\
$^{26}$IFPU -- Institute for Fundamental Physics of the Universe, Via Beirut 2, I-34014 Trieste, Italy \\
$^{27}$Department of Physics, Stanford University, Stanford, CA, USA 94305-4085 \\
$^{28}$Kavli Institute for Particle Astrophysics and Cosmology, 382 Via Pueblo Mall, Stanford, CA 94305-4060, USA \\
$^{29}$SLAC National Accelerator Laboratory, 2575 Sand Hill Road, Menlo Park, CA 94025, USA \\
$^{30}$Max-Planck-Institut f\"ur extraterrestrische Physik, Gie\ss enbachstra\ss e 1, 85748 Garching, Germany \\
$^{31}$Dipartimento di Fisica, Sezione di Astronomia, Universit\`a di Trieste, Via G. B. Tiepolo 11, 34143 Trieste, Italy \\
$^{32}$INFN -- Sezione di Trieste, Via Valerio 2, 34127 Trieste, Italy \\
$^{33}$ICSC -- Centro Nazionale di Ricerca in High Performance Computing, Big Data e Quantum Computing, Bologna, Italy \\
$^{34}$Instituto de F\'isica, Pontificia Universidad Cat\'olica de Valpara\'iso, Casilla 4059, Valpara\'iso, Chile \\
$^{35}$Department of Physics, Yamagata University, 1-4-12 Kojirakawa-machi, Yamagata, Yamagata 990-8560, Japan \\
$^{36}$NASA/Goddard Space Flight Center, Greenbelt, MD, USA 20771 \\
}}          

\date{Accepted XXX. Received YYY; in original form ZZZ}

\pubyear{\the\year{}}

\begin{document} 
    \label{firstpage}
    \pagerange{\pageref{firstpage}--\pageref{lastpage}}
    \maketitle
    \begin{abstract}
        Measurements of the entropy and temperature distribution of the intracluster medium (ICM) of local clusters have proven pivotal for the study of cluster astrophysics. 
        Here, we extend these measurements to $z=1.71$ with a resolved analysis of the ICM in ACT/SPT-CL~J0459.6-4946. 
        By combining high-resolution thermal Sunyaev-Zeldovich measurements using new, deep ALMA Band 1 and 3 observations, with archival X-ray data, we conclude that the cluster has 
        (i) an ICM with a flat and relatively high ($\approx 200~$keV cm$^2$) entropy distribution out to $\approx0.3\,r_{\rm 500,c}$, 
        (ii) a hot core ($T_{\rm e}\simeq14$ keV) with a declining temperature profile, 
        (iii) a short effective path length along the line of sight of $l_{\rm eff}\simeq\left(0.75\pm0.05\right)\times r_{\rm 500,c}$, and 
        (iv) a clumpiness factor $C= \left<n^2_\text{e}\right>/\left<n_\text{e}\right>^2\simeq1.17\pm0.05$ within $r_{\rm 500,c}$. 
        Its dynamic nature and high redshift suggest that the cluster is still thermalising its ICM, consistent with a relatively large hydrostatic mass bias of $b = 1 - M_{\rm 500,c, HE}/M_{\rm 500,c, WL} = 0.27^{+0.32}_{-0.67}$ that we measure with respect to the weak lensing mass. 
        Furthermore, a region with low entropy and high density to the south of the cluster centre is observed in the X-rays yet absent in the Compton-$y$ map, indicating pressure equilibrium; we interpret this as a sloshing cold front --- the first such detection at high redshifts. 
        This case study demonstrates that $\approx3$h-long ALMA observations, combined with comparatively ``short'' X-ray exposures ($\approx122$\,ks), can fully characterise the ICM at high redshift, opening a path towards the thermodynamics of young and distant galaxy clusters for larger samples.
    \end{abstract}


\section{Introduction}
    
    Since \citet{Zwicky1933}'s work, observations of individual clusters of galaxies have been pivotal for the study of cosmology and for understanding the physics of the most massive overdensities formed in the Universe. It nevertheless took three more decades to detect \citep{Byram1966,Gursky1971}, and another to characterise \citep{Mitchell1976,Serlemitsos1977}, the hot intracluster medium (ICM), in which most of the clusters' baryonic mass resides. Since then, measurements of the ICM's thermodynamic properties have informed research on topics such as feedback by active galactic nuclei (AGN), low-density plasma physics, and cosmology \citep[for reviews, see;][]{Sarazin1986,Voit2005, Pratt2019}.
    
    These observations of the ICM, however, have largely been limited to the nearby Universe ($z<0.5$), mainly because the observed X-ray surface brightness suffers from cosmological dimming. This has not halted attempts to characterise the thermodynamic distribution of the ICM at early epochs ($z>1.5$; see, for instance, \citealt{McDonald2017,Mantz2020,Tozzi2022,Mantz2026}), but such observations require \textit{Chandra} and/or \textit{XMM-Newton} integration times approaching the megasecond regime. These long integrations are essential for obtaining spectroscopic temperature estimates of the hot electrons in the ICM by fitting a plasma model to the observed X-ray spectrum. The temperature estimates, in turn, are crucial for accurately inferring cosmological parameters \citep{Pratt2019} and for studying astrophysical phenomenon in clusters, such as shock fronts, cooling flows, and sloshing motions \citep[e.g.,][]{Markevitch2007,McNamara2012}.    
    
    \begin{figure}
        \centering
        \includegraphics[width=0.9\hsize]{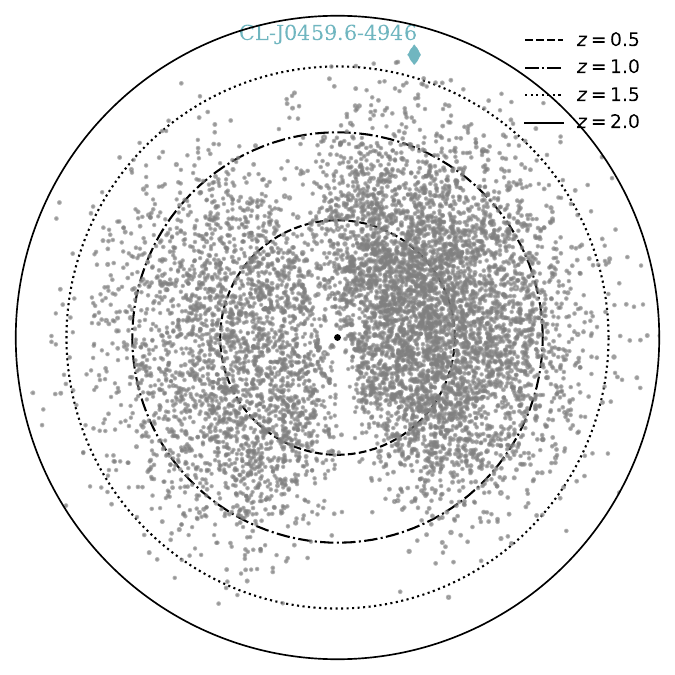}
        \caption{
        Wedge plot of the ACT DR6 cluster catalogue for clusters with S/N $\geq 5.5$ (gray points), drawn in the equatorial plane. The marker sizes scale with cluster mass ($\rm M_{500,c}E(z)$). RA is used as the angular coordinate, while the radial coordinate represents the comoving distance computed from their respective redshifts. Circles mark the distances corresponding to $z=0.5,~ 1.0,~1.5~\text{and},2.0$. We highlight the location of CL~J0459 (blue diamond-shaped marker). 
        }
        \label{fig:sample}
    \end{figure}
    
    A much more time-efficient approach to obtaining estimates of thermodynamic properties of the ICM is to use observations in both the millimetre (mm) and X-ray regimes (see, e.g., \citealt{Birkinshaw1991}, \citealt{Hughes1998}, \citealt{Mroczkowski2009, Mroczkowski2012}, \citealt{Adam2017}, \citealt{Siegel2018}, \citealt{Shitanishi2018}, \citealt{Castagna2020}, \citealt{Churazov2021}, \citealt{Ruppin2021}, \citealt{Okabe2021}, \citealt{Andreon2023}, and \citealt{Lepore2024}). The mm-wave regime probes the hot ICM through inverse Compton scattering of cosmic microwave background (CMB) photons, a phenomenon known as the thermal Sunyaev-Zeldovich effect (tSZ effect; \citealt{SZ1972}; for a review of the tSZ effect and its applications in astrophysical studies, see \citealt{Mroczkowski2019}). Because the tSZ effect probes the integrated thermal electron pressure along the line of sight, and the surface brightness of the X-ray emission is roughly proportional to the line-of-sight integral of the square of the thermal electron density, one can infer a temperature map of the ICM using the ideal gas law. Therefore, the ratio of the tSZ flux to the square root of the X-ray flux yields a pseudo, electron-density-weighted temperature distribution at a much higher angular resolution than usually achievable when deploying the X-ray-only spectroscopic method \citep[see also,][]{Wicker2026}.

    In this work, we derive high-resolution ($\approx6\arcsec$) thermodynamic distributions of the electron density, pressure, temperature, and entropy of a young galaxy cluster at $z=1.7$, known as ACT/SPT-CL J0459.6-4946 (hereafter CL~J0459; see Figure~\ref{fig:sample}, RA=$74.9230\,^\circ$, Dec =$ -49.7829\,^\circ$). Since its discovery, CL~J0459 has attracted a fair amount of attention \citep[see also][]{McDonald2017, Strazzullo2019, Mantz2020, Flores2021, Ghirardini2021, Zohren2022, Strazzullo2023, Mantz2026} as the highest-redshift object found in the wide-field South Pole Telescope (SPT) SZ survey \citep{Bleem2015}. Meanwhile, another CMB survey telescope, the Atacama Cosmology Telescope (ACT; \citealt{Hilton2021}), also detected the object and has since identified similar sources at comparably high redshifts \citep{Hilton2026}. The same holds for the latest SPT data releases \citep{Bleem2024, Kornoelje2026}. Nevertheless, CL~J0459 is only the second cluster at $z>1.5$ for which deep, resolved observations across the spectrum are currently available, enabling studies of 
    its thermodynamic properties (the other being XLSSC~122 at $z=1.98$; \citealt{vanMarrewijk2024}). 
    
    Here we present new, deep Atacama Large Millimetre/submillimeter Array (ALMA) observations of the tSZ effect towards CL~J0459, resolving the electron pressure distribution of the ICM at arcsecond scales. We combine these with ACT data and archival X-ray observations from \textit{Chandra} \citep{McDonald2017} and XMM-Newton \citep{Mantz2026} to map the thermodynamic properties of this young galaxy cluster. 

    The remainder of this work is structured as follows. Section~\ref{sec:observations} describes the data used in the analysis. Section~\ref{sec:galaxies} presents the compact line- and continuum-emitting sources detected in the mm-wave data. Section~\ref{sec:dynamics} describes the analysis of the thermalisation state of the ICM, and Section~\ref{sec:discussion} discusses its astrophysical implications. Finally, we summarize our conclusions in Section~\ref{sec:conclusions}. For all calculations, the assumed cosmology is a spatially flat, $\Lambda$ cold dark matter ($\Lambda$CDM) model with $H_0=70$ km\,s$^{-1}$Mpc$^{-1}$ and $\Omega_{\rm m} = 0.3$. At the cluster redshift ($z=1.71$) the physical scale is $1\arcsec = 8.463$\,kpc. We note that at these redshifts differences in the angular diameter distance (DA) between different cosmologies, for instance \citet{Riess2022} and \citet{Planck2020}, are on the order of $8\%$. 
    
\section{Observations}\label{sec:observations}

    \begin{figure}
        \centering
        \includegraphics[width=\hsize]{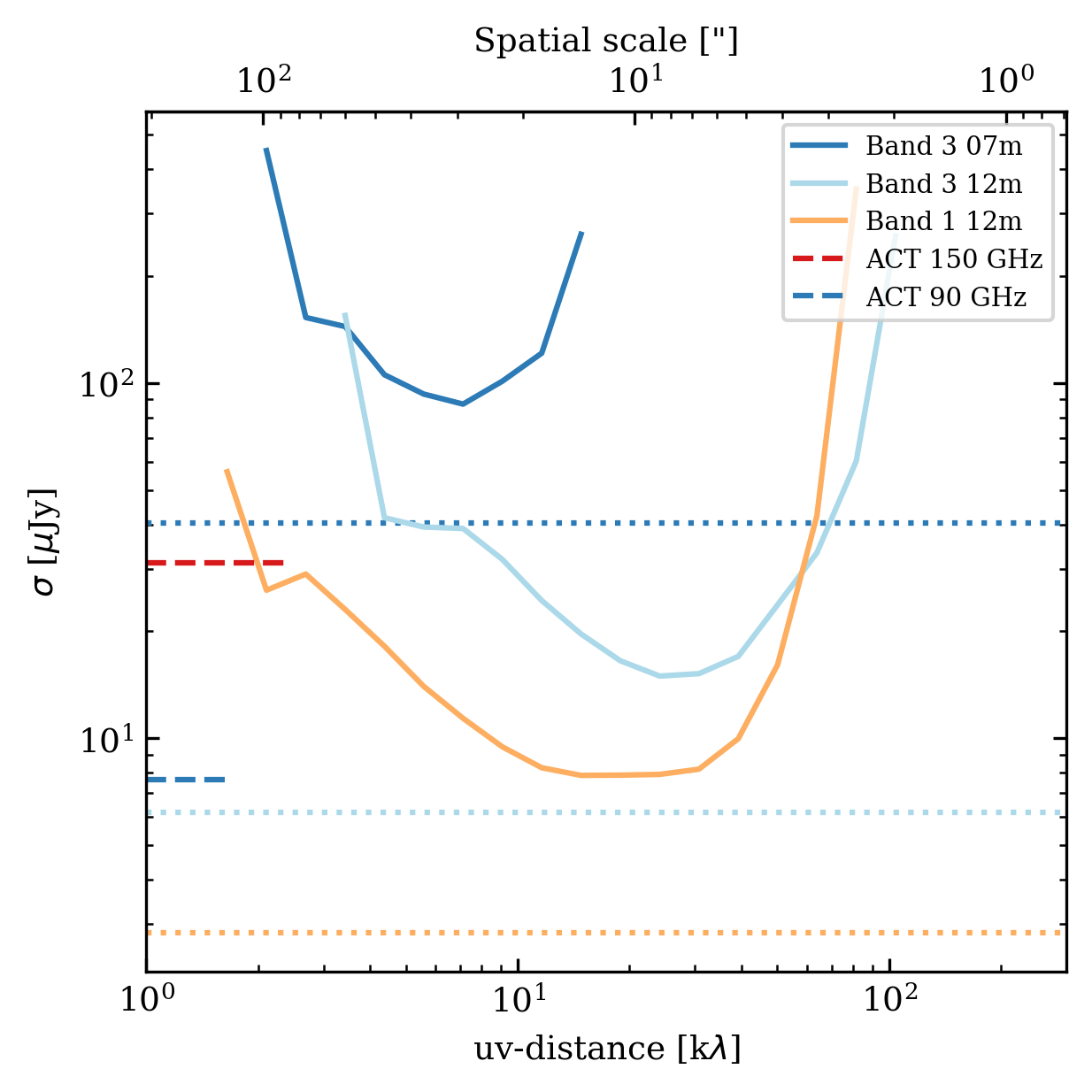}
        \caption{Sensitivity per binned Fourier mode sampled for the various ALMA and ACT observations of CL~J0459 (not corrected for the different transfer functions). The horizontal dotted lines indicate the various white noise levels of the observations, measured in the visibility plane for the central pointing. The dashed lines, up to the FWHM of their respective beam size, show the point source sensitivity of the ACT DR6 data at the location of CL~J0459.}
        \label{fig:sensitivity}
    \end{figure}

    \subsection{Millimetre-wave observations}

        To map the tSZ effect in the mm-wave regime, we make use of both ACT and ALMA observations. Together, these probe the pressure distribution over a broad range of spatial scales, with ACT functioning as the zeroth baseline, anchoring the large-scale pressure distribution that is resolved out by the interferometer, while the ALMA visibilities constrain the higher angular modes. The interferometric data are described in Section~\ref{sec:ALMA}, while Section~\ref{sec:ACT} briefly describes the ACT observations.    
        
        \subsubsection{ALMA}\label{sec:ALMA}
        
            In this work, we present ALMA data on cluster CL~J0459 (Project IDs: 2022.1.00427.S, 2023.1.00213.S, 2023.1.01638.S; PI: T. Kitayama, F. Gómez; Bands 1 \& 3). All calibrated measurement sets are obtained via the \texttt{calMS} service \citep{Petry2020} of the European ALMA Regional Centre \citep{Hatziminaoglou2015}. All post-processing and imaging steps are performed using the latest version of \texttt{casatasks} available at the time of the analysis (v6.6.5.31). No additional flagging or recalibration is necessary. An observing problem in the corresponding cycle with ALMA Band 1 observations, however, is that they are affected by a recently detected Band 1 local oscillator (LO) leakage, which compromises the image quality of the affected spectral window \citep{Remijan2019}. The spike, located around 43.9~GHz, is visible in the quality assurance portion of the calibration pipeline, and hence, affected channels were removed from this dataset after calibration (see Appendix~\ref{app:LO}). The total on-source time is 30\,h for the Atacama Compact Array (ACA) Band 3 (7 pointings), 12\,h for 12m-array Band 3 (7 pointings), and 3\,h for 12m-array Band 1 (single pointing). Observing specifics are summarized in Table~\ref{tab:alma_data}

            \begin{table}
            	\centering
            	\caption{Summary of the ALMA observations of CL~J0459.6--4946.}
            	\label{tab:alma_data}
            	\footnotesize
            	\setlength{\tabcolsep}{1.4pt}
                \resizebox{0.48\textwidth}{!}{%
            	\begin{tabular}{lccc}
            		\toprule
            		Band & 1 & 3 & 3 \\
            		Array & 12-m & 12-m & 7-m (ACA) \\
            		\midrule
            		Project code & 2023.1.01638.S & 2023.1.00213.S & 2022.1.00427.S \\
            		PI & F.~G\'omez & T.~Kitayama & T.~Kitayama \\
            		Frequency (GHz) & 38.7--44.9 & 85.0--99.0 & 85.0--99.0 \\
                    Resolution (km~s$^{-1}$)$^a$ & 60 & 60 & 60 \\
            		Pointings & 1 & 7 & 7 \\
            		$t_{\rmn{int}}$ (h)$^{b}$ & 3.3 & 1.8 & 6.0 \\
            		$\sigma_{\rmn{N}}$ ($\umu$Jy\,beam$^{-1}$)$^{c}$ & 2.8 & 6.2 & 40.4 \\
            		Spatial Scales (\arcsec) & 3--135 & 2--125 & 14--134 \\
            		\bottomrule
            	\end{tabular}}
            	\begin{minipage}{\columnwidth}
            		\textbf{Notes}. 
                    $^a$ Native spectral resolutions differed for all observations and spectral windows within each observation and were thus imaged consistently with a width of 60 km/s. 
                    $^{b}$~On-source integration time per pointing. 
            		$^{c}$~White-noise level per pointing, derived from the visibility
            		weights. \\
            	\end{minipage}
            \end{table}
        
            \begin{figure}
                \centering
                \includegraphics[width=\linewidth]{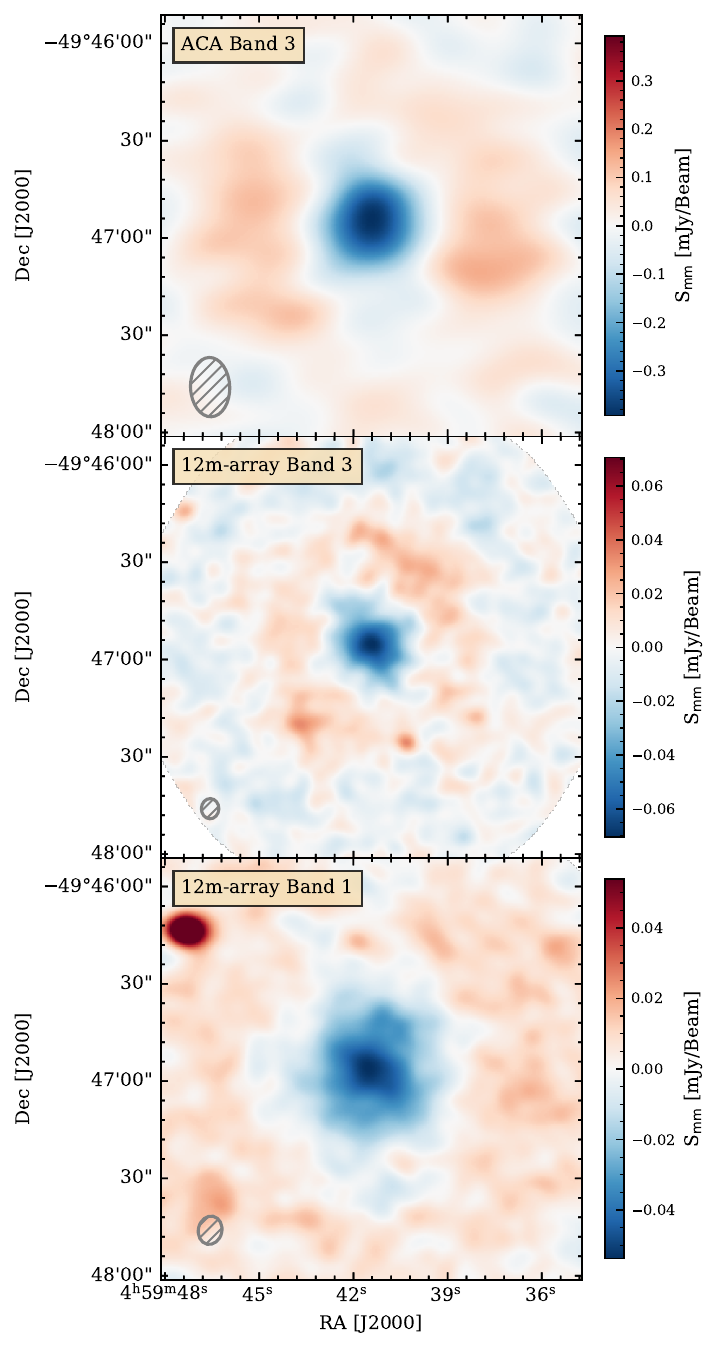}
                \caption{Dirty continuum maps of the three ALMA observations. From top to bottom, the three maps show respectively ACA Band 3, 12m-array Band~3, and 12m-array Band~1 data. The figure shows a clear detection of the tSZ flux at the centre with side lobes surrounding it.}
                \label{fig:dirty_maps}
            \end{figure}
    
            Figure~\ref{fig:sensitivity} shows how the various ALMA observations compare in terms of sensitivity per Fourier mode sampled. These specific ALMA Band 1 12m-array observations probe similar scales to those probed by the ACA in Band 3, but with nearly 10 times the point source sensitivity, and without the need to mosaic due to its larger field of view. We note that the increase in sensitivity does not translate linearly to higher S/N, as the tSZ flux is brighter at 90\,GHz than at 40\,GHz. For comparison, the dirty maps of the individual ALMA observations are shown in Figure~\ref{fig:dirty_maps}. They have been imaged using a natural weighting and mosaic gridding, as will all ALMA maps in this work.

    \subsubsection{ACT}\label{sec:ACT}

            The Atacama Cosmology Telescope (ACT) was a 6-meter, off-axis Gregorian survey telescope that operated from 2007 to 2022 in the Atacama Desert of Chile \citep{Swetz_2011, Thornton_2016, henderson2016advanced}. Over its lifetime, the detectors and receiver were modified such that it covered, at various times, frequency bands centred at approximately 30, 40, 90, 150, 220, and 280\,GHz. In this work, we use the bands centred at the frequencies most sensitive to the tSZ effect, the 90 and 150\,GHz, which were obtained with the AdvACT receiver that operated from 2017 onward. At these frequencies, the ACT data have a FWHM resolution of $\approx2\arcmin$ and $\approx1.4\arcmin$, respectively. The noise levels around CL~J0459 are shown in Figure~\ref{fig:sensitivity}.
    
    \subsection{X-ray observations}\label{sec:X-ray_data}
    
        \begin{figure}
            \centering
            \includegraphics[width=\linewidth]{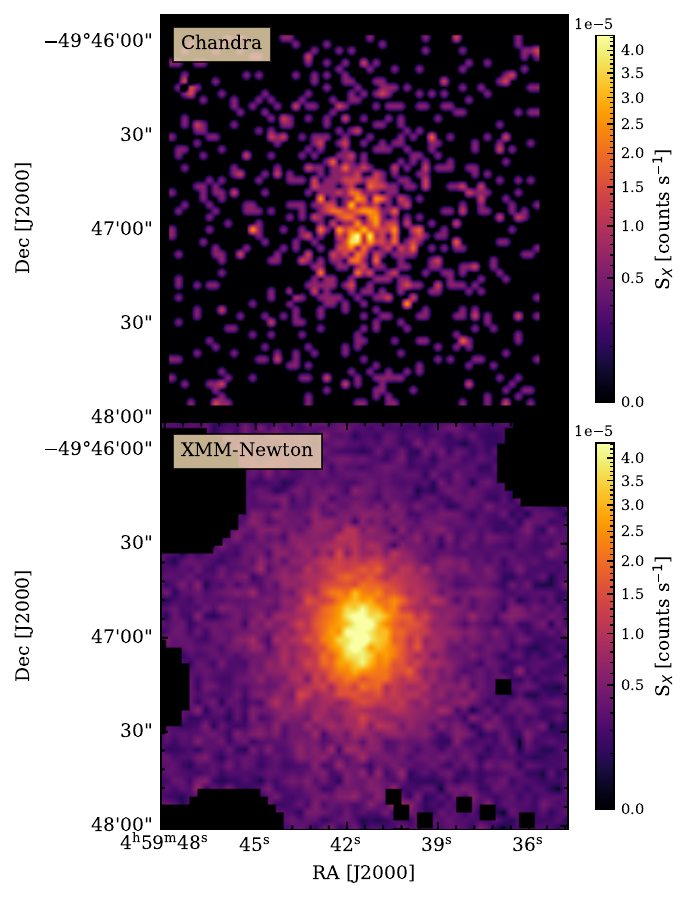}
            \caption{Broadband X-ray surface brightness maps of CL~J0459 obtained with \textit{Chandra} and XMM-Newton plotted within the same FoV. The data were first reported in \citet{McDonald2017} and \citet{Mantz2020}, respectively. Contaminating point sources have been masked and removed. The \textit{Chandra} data are integrated over the energy range $0.5-2$~keV, while the XMM-Newton data are integrated over $0.4-4$~keV. The figures use a square-root scaling.}
            \label{fig:X-rays}
        \end{figure}
    
        We make further use of the archival XMM-Newton X-ray observations with 699~ks of clean on-source time \citep[][PI; A.\ Mantz]{Mantz2020,Mantz2026} and archival \textit{Chandra} data \citep[][PI; M.\ McDonald, 122~ks clean exposure time]{McDonald2017}. The XMM-Newton data, constrained via spectral modelling of the (rest-frame) 6.7~keV Fe emission line complex, provided global ICM properties of $kT=7.2\pm0.3$~keV, $Z/Z_\odot = 0.37\pm0.08$, and $z = 1.71 \pm 0.02$ within $60\arcsec$ (corresponding to 510~kpc, \citealt{Mantz2020,Flores2021, Mantz2026}). Here, the sensitivity of both the \textit{Chandra} and XMM-Newton data was insufficient for measuring a radial temperature and entropy distribution beyond an isothermal profile.\footnote{\citet{Mantz2020}'s and \citet{Mantz2026}'s spectral analysis was limited by the lack of surface brightness to two bins, which were statistically consistent with an isothermal temperature profile.}
        
        There are six \textit{Chandra} ACIS-I observations of this cluster (ObsIDs: 17211, 17501, 17502, 18711, 18824, 18853). The \textit{Chandra} observations were each reprocessed using \texttt{CIAO 4.18.0} with Calibration Database \texttt{CALDB 4.12.0}. After light curve filtering to remove any potentially flared data, the observations were merged and exposure corrected using the \texttt{merge\_obs} task, binning the native 0$\farcs$492 pixels $8\times8$, resulting in combined maps with 1$\farcs$968 pixels to reduce the effect of PSF smearing.

        For the XMM-Newton surface brightness map, we use the exposure-corrected, merged EPIC MOS1+MOS2+PN flux map processed with \texttt{XMM-SAS v16.1.0}, integrating over the $0.4$--$4$~keV energy range. Point sources identified via \texttt{CIAO} tool \texttt{wavdetect} from the \textit{Chandra} data were masked prior to analysis in XMM-Newton. The respective broadband XMM-Newton and \textit{Chandra} maps of CL~J0459 are shown in Figure~\ref{fig:X-rays}. A list containing the X-ray data used can also be found in Table~4 of \citet{Mantz2026}. 

\section{Cluster members seen in the mm-wave regime}\label{sec:galaxies}

    With ALMA's capability of detecting both cold gas in cluster members via line and continuum emission and hot gas around the galaxies through the tSZ effect, we need to accurately characterise both to correct for contaminants when studying one or the other. In this work, we consider the cluster members to be contaminants to the tSZ effect. Hence, in this section we report the detected line emission (Section~\ref{sec:line_sources}) and continuum emission of compact sources in the field (Section~\ref{sec:cont_sources}), without going in further detail on their astrophysical implications.
    
    \begin{figure}
        \centering
        \includegraphics[width=\linewidth]{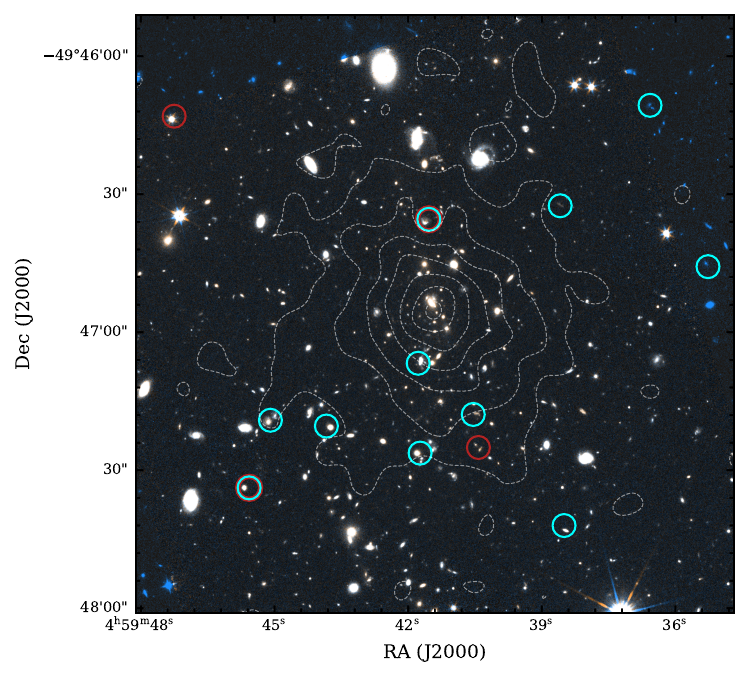}
        \caption{HST background (F140W/F110W) of CL~J0459 for which we highlight the position of bright lines in the ALMA Band 3 data (cyan) and Band 1 (red). These lines are also clearly evident in the moment-8 map shown in Appendix~\ref{fig:moment8_} and average to a redshift of $z=1.71\pm0.02$. White contours show the deconvolved tSZ data at $\left[-26, -22, -18, -14, -10, -6, -2\right]-\sigma$ from Figure~\ref{fig:JvM_clean}.}
        \label{fig:sources}
    \end{figure}
    
    \subsection{Line-emitting sources}\label{sec:line_sources}
    
        At $z=1.71$ \citep{Mantz2020}, the CO~J=1--0 and CO~J=2--1 transitions of CL~J0459 are redshifted into ALMA Bands~1 and~3, respectively, and the spectral windows were configured to cover them. We detect four line-emitting sources in the Band~1 data and eleven in Band~3, all at $S/N>4$ when integrated over the full linewidth. All detected line emitters are listed in Table~\ref{tab:lines}, and location of the lines are shown in Figure~\ref{fig:sources} overlaid on a colour composite image of the cluster using Hubble Space Telescope (HST) data. 
        
        To retrieve fluxes and other line properties, we first performed a continuum subtraction using \textsc{CASA}-\texttt{uvcontsub} with \texttt{fitorder}=1 on all channels in the four basebands individually for the ACA and 12m-array. Then, we jointly \texttt{clean}ed the 12m-array data with the ACA using manual spatial-kinematic masking and imaged them using a 60~km\,s$^{-1}$ channel width. Any bright lines present are easily identified by eye using the moment-8 maps (See Appendix~\ref{app:members}). Any faint lines possibly present in the data will not bias the tSZ flux characterisation when averaged over the full spectral window \citep[see also the discussion in ][]{Kitayama2023}. The bright lines, however, might contaminate continuum maps and bias the outcome. Therefore, we used the \texttt{clean} model to subtract the bright line emission from the visibilities using \textsc{CASA} tasks \texttt{ft} and \texttt{uvsub}, before averaging each spectral window into a single frequency bin.\footnote{This last step was necessary to reduce the data volume before forward-modelling parametric profiles to the visibilities (see section~\ref{sec:profiles}).} The \textsc{casa} task \texttt{ft} makes a visibility based model of a data cube, matching the $uv$-coverage of the observations, while \texttt{uvsub} subtracts that model from the data.
        
        Two line emitters found in Band 1 have no line detection at the same location in Band 3 (see red circles in Fig~\ref{fig:sources}), while they match continuum sources detected in both Band 1 and 3. These two line emitters are cross-matched with WISE and Radio sources in NED (NASA/IPAC Extragalactic Database), and none have previously reported redshifts.
        Expecting the CO J=2-1 line to be at least twice as bright as the CO J=1-0 transition \citep{Carilli2013}, and given that our Band 1 data has a white noise level of $0.48\times$ that of Band 3 at the phase centre, we would have detected a CO J=2-1 line in the Band 3 data given the sensitivity levels at the same significance. Therefore, we assume that lines detected only in Band 1 are foreground or background sources rather than cluster members. For the remaining 11 sources detected in Band 3 we find that two of them have matching line emission of what would be the CO J=1-0 line. 
        
        We measure the redshift of each Band 3 line emitter by fitting a Gaussian to its extracted spectrum (Appendix~\ref{app:members}). The 11 emitters have a median redshift of $z=1.717\pm0.011$, with the uncertainty from bootstrap resampling, which is consistent with the ICM redshift of $z=1.705\pm0.018$ that \citet{Mantz2020} derived from X-ray spectra of the ICM. We note, however, that their velocity dispersion along the line of sight, estimated with the gapper method \citep{Beers1990}, is $2320\pm320$\,km\,s$^{-1}$. This high value suggests that the emitters are unlikely to all be virialised members, and thus may trace structures around the cluster along the line of sight outside of the virial regime. 
        
    \subsection{Continuum sources}\label{sec:cont_sources}
    
        After identifying and removing line emission from the visibilities, we searched for continuum sources using baselines longer than $10\,\mathrm{k}\lambda$ to filter out the extended tSZ flux. The line-emitting and continuum sources are treated as independent contaminants. We found five continuum sources in the Band 3 ALMA data, two of which are also detected in Band 1. The fluxes are measured by forward modelling the extended tSZ surface brightness distribution jointly with the point sources. The resulting measured positions, fluxes, and spectral slopes are listed in the Appendix, Table~\ref{tab:cont_fluxes}. For the remainder of the work, any point-source emission---both lines and continuum---is removed from the ALMA data before plotting, both in the image- and $uv$-plane.
        
\section{ICM dynamics}\label{sec:dynamics}

    The primary goal of this work is to combine archival X-ray data with new, high-resolution ALMA and low resolution ACT observations to constrain the thermal pressure, density, temperature, and entropy distributions of the ICM at an angular resolution of $\approx6\arcsec$ ($\approx50\,$kpc). To this end, we adopt two complementary, linked approaches: 
    (i) In a forward model approach, we infer radial thermodynamic profiles of the ICM (see Section~\ref{sec:profiles}) and compare our results with the outcome of the analysis done by \citet{Mantz2026} on the $\sim6\times$ deeper XMM-Newton data than the \textit{Chandra} data we use. 
    (ii) In an inverse approach, we construct line-of-sight-projected maps of all thermodynamic quantities to search for asymmetries in the 2D distributions (see Section~\ref{sec:maps}) going beyond the symmetric approach presented in Section~\ref{sec:profiles}. These two methods are intrinsically connected, though: the maps inform the choice of parametrisation used in the modelling, while the profile modelling is employed, for example, to deconvolve the ALMA data, which in turn is used to compute the respective maps. Although presented in separate subsections, both approaches are cross-referenced throughout the rest of this section.

    \subsection{Fitting parametric profiles} \label{sec:profiles}
        
        We adopt a forward-modelling approach to infer the radial thermodynamic profiles of CL~J0459. We first model the thermal pressure distribution using the ALMA and ACT data following \citet{vanMarrewijk2024}, as outlined in Section~\ref{sec:fit_ALMA}, which is based on the method of \citet{DiMascolo2019}. The resulting maximum-posterior model is then used in the X-ray analysis to jointly constrain the radial density, temperature, and entropy profiles, employing the \texttt{MBProj2D} code developed by \citet{Bulbul2024} and \citet{Sanders2026} with an adaptation similar to that of \citet{Scheck2023}, as described in Section~\ref{sec:fit_Chandra}. Both implementations share the same underlying principles: parametric modelling of radial profiles with likelihood evaluation performed directly in the native two-dimensional observational plane, while allowing for non-spherical morphologies. Furthermore, by jointly modelling the ICM and point source contaminants, we can reliably isolate point source emission without over-subtracting ICM flux --- a step that is crucial for interpreting the thermodynamic maps (see Section~\ref{sec:maps}).
        
        \subsubsection{The radial pressure profile}\label{sec:fit_ALMA}

            \begin{table}
            	\centering
                \caption{Priors adopted for the tSZ profile modelling.}
            	\label{tab:priors_tSZ} 
            	\footnotesize
            	\setlength{\tabcolsep}{3pt}
                \begin{tabular}{lll}
                    \toprule
                     & A10 & gNFW \\
                    Parameter &  &  \\
                    \midrule
                    $\Delta$RA ["] & $\mathcal{U}(-30, 30)$ & $\mathcal{U}(-30, 30)$ \\
                    $\Delta$Dec ["] & $\mathcal{U}(-30, 30)$ & $\mathcal{U}(-30, 30)$ \\
                    $M_{\rm 500,c}$ [$10^{14}$M$_\odot$] & log-$\mathcal{U}(1, 5)$ & — \\
                    $p_\mathrm{norm}^{\dagger}$ & $8.4$ & log-$\mathcal{U}(0.001, 1)$ \\
                    $c_{500}$ & $1.18$ & — \\
                    $r_s$ ["] & — & log-$\mathcal{U}(1.18, 360)$ \\
                    $e$ & $\mathcal{U}(0, 0.9)$ & $\mathcal{U}(0, 0.9)$ \\
                    PA [deg] & $\mathcal{U}(-90, 90)$ & $\mathcal{U}(-90, 90)$ \\
                    $\alpha$ & $1.05$ & $\mathcal{U}(0.5, 10)$ \\
                    $\beta$ & $5.49$ & $\mathcal{U}(3, 10)$ \\
                    $\gamma$ & $0.308$ & $\mathcal{U}(0, 0.99)$ \\
                    $\alpha_\mathrm{0}$ & $0.22$ & — \\
                    $z$ & $\mathcal{N}(1.71, 0.02)$ & $\mathcal{N}(1.71, 0.02)$ \\
                    $f_\mathrm{cal}$ & $\mathcal{N}(1, 0.05)$ & $\mathcal{N}(1, 0.05)$ \\
                    \bottomrule
                \end{tabular}
                \begin{minipage}{\columnwidth}
                \textbf{Notes}. $\Delta$RA and $\Delta$Dec are offsets from $(\rmn{RA},\rmn{Dec}) = (74\fdg92296, -49\fdg78184)$. $f_{\rmn{cal}}$ is the flux-calibration factor, one per data set. The redshift prior is centred on the X-ray spectroscopic redshift. Both are used to marginalise over these parameters, rather than constrain them. $^{\dagger}$ Units of $p_{\rm norm}$ are given in keV\,cm$^{-3}$ for the gNFW runs and are dimensionless in the case of the A10-run. 
               \end{minipage}
            \end{table}

            \begin{table*}
                \caption{CL~J0459 pressure profile parameters for various model runs} 
                \label{tab:params}
                \setlength{\tabcolsep}{4pt}
                \begin{threeparttable}
                \resizebox{\textwidth}{!}{%
                \begin{tabular}{lllllllllllll}
                \toprule
                                                        & $\Delta$RA [\arcsec]    & $\Delta$Dec [\arcsec]  & $M_{\rm 500,c}$                  & $p_\mathrm{norm}$            & $c_{500}$ & $r_{\rm s}$ [\arcsec] & $e$                    & $\mathrm{PA}$  & $\alpha$               & $\beta$             & $\gamma$               & $\Delta\ln\mathcal{Z}^\dagger\ (\sigma)$ \\
                Model                                   &                         &                        & [$\times10^{14}\,{\rm M_\odot}$] &                              &           &                       &                        &                &                        &                     &                        &                                          \\
                \midrule
                A10 (sph)                               & $-0.28^{+0.24}_{-0.24}$ & $1.49^{+0.25}_{-0.25}$ & $2.23^{+0.27}_{-0.24}$           & $8.4$                        & $1.18$    & —                     & $0.00$                 & $0$            & $1.05$                 & $5.49$              & $0.308$                & $2397.3\ (69.2\sigma)$                   \\
                A10 (ell)                               & $-0.25^{+0.23}_{-0.23}$ & $1.49^{+0.26}_{-0.26}$ & $2.42^{+0.30}_{-0.26}$           & $8.4$                        & $1.18$    & —                     & $0.14^{+0.02}_{-0.02}$ & $14^{+5}_{-5}$ & $1.05$                 & $5.49$              & $0.308$                & $+9.1\ (+4.3\sigma)$                     \\
                gNFW A10 $\gamma$-free (sph)            & $-0.30^{+0.24}_{-0.25}$ & $1.51^{+0.26}_{-0.26}$ & —                                & $0.093^{+0.046}_{-0.030}$    & —         & $63^{+10}_{-9}$       & $0.00$                 & $0$            & $1.05$                 & $5.49$              & $0.33^{+0.12}_{-0.13}$ & $+4.8\ (+3.1\sigma)$                     \\
                gNFW A10 $\gamma$-free (ell)            & $-0.24^{+0.24}_{-0.24}$ & $1.53^{+0.28}_{-0.28}$ & —                                & $0.075^{+0.036}_{-0.024}$    & —         & $76^{+12}_{-11}$      & $0.17^{+0.02}_{-0.02}$ & $14^{+4}_{-4}$ & $1.05$                 & $5.49$              & $0.39^{+0.11}_{-0.12}$ & $+23.2\ (+6.8\sigma)$                    \\
                gNFW M14 $\gamma$-free (sph)            & $-0.29^{+0.25}_{-0.25}$ & $1.52^{+0.26}_{-0.26}$ & —                                & $0.041^{+0.007}_{-0.006}$    & —         & $30.8^{+2.3}_{-2.4}$  & $0.00$                 & $0$            & $2.27$                 & $3.48$              & $0.73^{+0.06}_{-0.08}$ & $+3.1\ (+2.5\sigma)$                     \\
                gNFW M14 $\gamma$-free (ell)            & $-0.23^{+0.24}_{-0.24}$ & $1.53^{+0.28}_{-0.27}$ & —                                & $0.036^{+0.007}_{-0.006}$    & —         & $36^{+4}_{-3}$        & $0.17^{+0.02}_{-0.02}$ & $14^{+4}_{-4}$ & $2.27$                 & $3.48$              & $0.78^{+0.07}_{-0.08}$ & $+21.1\ (+6.5\sigma)$                    \\
                gNFW M14 $\beta\gamma$-free (sph)       & $-0.30^{+0.24}_{-0.25}$ & $1.52^{+0.26}_{-0.26}$ & —                                & $0.083^{+0.063}_{-0.034}$    & —         & $73^{+44}_{-31}$      & $0.00$                 & $0$            & $1.05$                 & $6.0^{+2.4}_{-1.6}$ & $0.35^{+0.12}_{-0.15}$ & $+4.4\ (+3.0\sigma)$                     \\
                gNFW M14 $\beta\gamma$-free (ell)       & $-0.23^{+0.24}_{-0.24}$ & $1.53^{+0.28}_{-0.28}$ & —                                & $0.065^{+0.051}_{-0.026}$    & —         & $91^{+52}_{-39}$      & $0.17^{+0.02}_{-0.02}$ & $14^{+4}_{-4}$ & $1.05$                 & $6.2^{+2.4}_{-1.8}$ & $0.42^{+0.11}_{-0.14}$ & $+22.9\ (+6.8\sigma)$                    \\
                gNFW M14 $\alpha\beta\gamma$-free (sph) & $-0.30^{+0.24}_{-0.24}$ & $1.52^{+0.25}_{-0.25}$ & —                                & $0.070^{+0.074}_{-0.031}$    & —         & $59^{+50}_{-27}$      & $0.00$                 & $0$            & $1.16^{+0.42}_{-0.23}$ & $5.3^{+2.7}_{-1.7}$ & $0.42^{+0.20}_{-0.21}$ & $+1.7\ (+1.9\sigma)$                     \\
                gNFW M14 $\alpha\beta\gamma$-free (ell) & $-0.23^{+0.24}_{-0.24}$ & $1.53^{+0.27}_{-0.26}$ & —                                & $0.061^{+0.070}_{-0.028}$    & —         & $77^{+62}_{-35}$      & $0.17^{+0.02}_{-0.02}$ & $14^{+4}_{-4}$ & $1.11^{+0.39}_{-0.21}$ & $5.6^{+2.7}_{-1.7}$ & $0.45^{+0.19}_{-0.21}$ & $+20.2\ (+6.4\sigma)$                    \\
                \bottomrule
                \end{tabular}}
                \end{threeparttable}
                \begin{minipage}{\textwidth}
                    \textbf{Notes.} $^\dagger$ Here, we quote the log difference between the null-test and the computed Bayesian evidence for the A10 (sph) run. The remaining $\Delta\ln\mathcal{Z}$ gives the log-evidence relative to the reference model A10 (sph), rather than the absolute value. Furthermore, halo masses are omitted from the gNFW-runs as it is not directly modelled. Furthermore, given that converting $Y_{\rm 500,c}$ to $M_{\rm 500,c}$ is accompanied with a plethora of assumptions, which might be less valid at high redshift, we opted to adopt only the A10-formalism to estimate the tSZ derived halo masses. The error on the halo mass is dominated by the intrinsic scatter of converting $Y_{\rm 500,c}$ to $M_{\rm 500,c}$ following \citet{Hilton2021}.
                \end{minipage}
            \end{table*}
            
            As pressure models, we consider the universal pressure profile (UPP) of \citet{Arnaud2010}, hereafter the A10--UPP, and the generalised NFW (gNFW) profile \citep{Nagai2007},
            \iftrue
            given by
            \begin{equation}\label{eq:gNFW}
                P_{\rm e}(r) = P_{\rm e,i}
                \left(\frac{r}{r_{\rm s}}\right)^{-\gamma}
                \left[1 + \left(\frac{r}{r_{\rm s}}\right)^{\alpha}\right]^{(\gamma - \beta)/\alpha},
            \end{equation}
            where $P_{\rm e,i}$ is the electron pressure normalisation and $\gamma$, $\alpha$, and $\beta$ are the shape parameters of the broken power law, describing the logarithmic slopes at small, intermediate, and large radii, respectively, relative to the scale radius $r_{\rm s}$. In this formulation, $P_{\rm e}$ depends on the radial distance $r$ from the cluster centroid, $(x_{\rm gNFW}, y_{\rm gNFW})$.
            
            To connect the gNFW profile to the self-similar A10--UPP formulation, we rewrite the radial coordinate as
            \begin{equation}
                \frac{r}{r_{\rm s}} = c_{500,\mathrm{c}}\,x,
            \end{equation}
            where $x = r / r_{500,\mathrm{c}}$ and $c_{500,\mathrm{c}} = r_{500,\mathrm{c}} / r_{\rm s}$ is the ICM concentration parameter.\footnote{$M_{500,\mathrm{c}}$ is defined as the total mass of the cluster within a radius $r_{500,\mathrm{c}}$ in which the average density is 500 times the critical density of the Universe at that redshift.} The self-similar pressure profile is then obtained using the local $M_{500,\mathrm{c}}$--$Y_{\rm X}$ relation by \citet{Arnaud2010} calibrated on the REXCESS sample, such that
            \begin{equation}
                P_{\rm self\text{-}similar}(M_{500,\mathrm{c}}, z)
                = P_{\rm e}(r)\,P_{500,\mathrm{c}}(M_{500,\mathrm{c}}, z),
            \end{equation}
            with $P_{\rm e}(r)$ given by Eq.~\eqref{eq:gNFW}. Following \citet{Arnaud2010}, the mass and redshift dependence enters through the normalisation term
            \begin{equation}
                \begin{split}
                    P_{500,\mathrm{c}}(M_{500,\mathrm{c}}, z) = 1.65 \times 10^{-3}\, E(z)^{8/3} \\
                    \times \left[\frac{M_{500,\mathrm{c}}}{3 \times 10^{14}\,\rm M_\odot}\right]^{\frac{2+ 3a_p(x)}{3}},
                \end{split}
            \end{equation}
            expressed in units of keV\,cm$^{-3}$. Here, $E(z) = H(z)/H_0$ is the dimensionless Hubble parameter, and $a_p(x)$ accounts for deviations from self-similarity in cluster cores, defined as
            \begin{equation}
                a_p(x) = \frac{a_0}{1 + 8\,x^3}~.
            \end{equation}
            When modelling the data via the A10 formalism, we set the shape parameters 
            to the A10--UPP values, with $a_0 = 0.22$ \citep[see discussion in][]{Arnaud2010}.
            \fi

            The profiles are projected onto a two-dimensional grid by integrating along the line of sight, and the likelihood is evaluated in the Fourier domain, with corrections for the transfer functions of the different ACT and ALMA datasets and the frequency-dependent conversion from integrated pressure to flux. The ACT transfer function consists of smoothing with the frequency-dependent beam and building a noise covariance model directly from the data following \citet{Gill2026}. The ALMA transfer function consists of multiplying the input map with the primary beam before Fourier transforming the map and interpolating  over the data-specific $uv$-coverage. This operation automatically corrects for the limited resolution of the observation.
            
            For the A10--UPP runs, we vary only the centroid position and halo mass to enable a direct mass comparison with the literature. For the gNFW model, all shape and position parameters are inferred simultaneously. For all inferences, we use nested sampling via the \texttt{nautilus} package \citep{Skilling2006, Lange2023}. 
            We further allow for ellipticity on the projected sky, with
            \begin{equation}
                e = 1-\frac{b}{a},
            \end{equation}
            with $b$ and $a$ the minor and major semi-axes in the plane of the sky, respectively. Since the extent along the line of sight is unconstrained by the projected signal, we fix the corresponding semi-axis to the inverse root mean square average of $a$ and $b$, following \citet{DiMascolo2019}. Note that this assumption impacts the derived halo mass through the effective path length along the line of sight \citep[see discussion in][]{DiMascolo2019}. All parameters with priors are listed in Table~\ref{tab:priors_tSZ}.

        \subsubsection{Interpretation of the mm-wave analysis}
    
            \begin{figure*}
                \centering
                \includegraphics[width=0.49\textwidth]{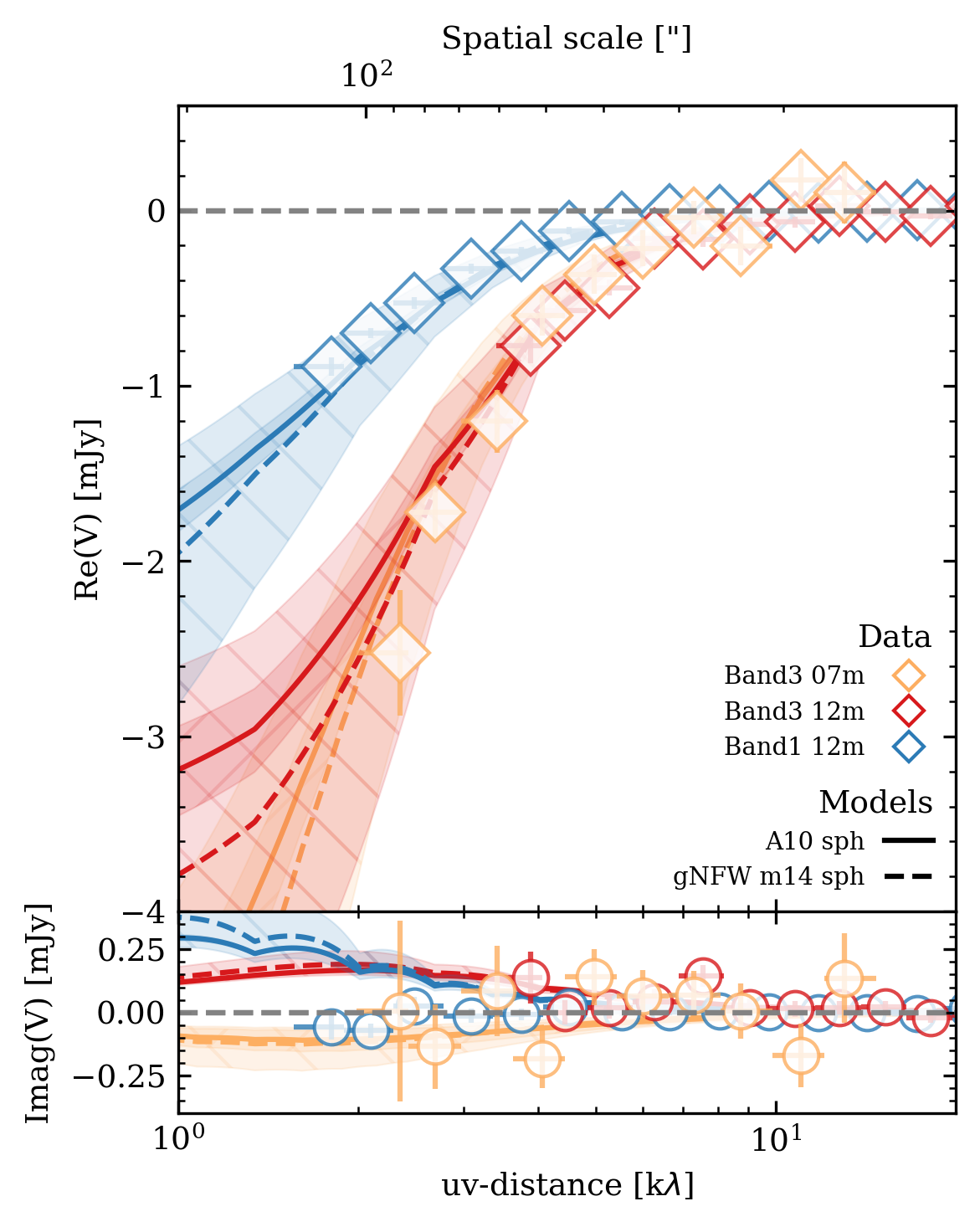}
                \includegraphics[width=0.49\textwidth]{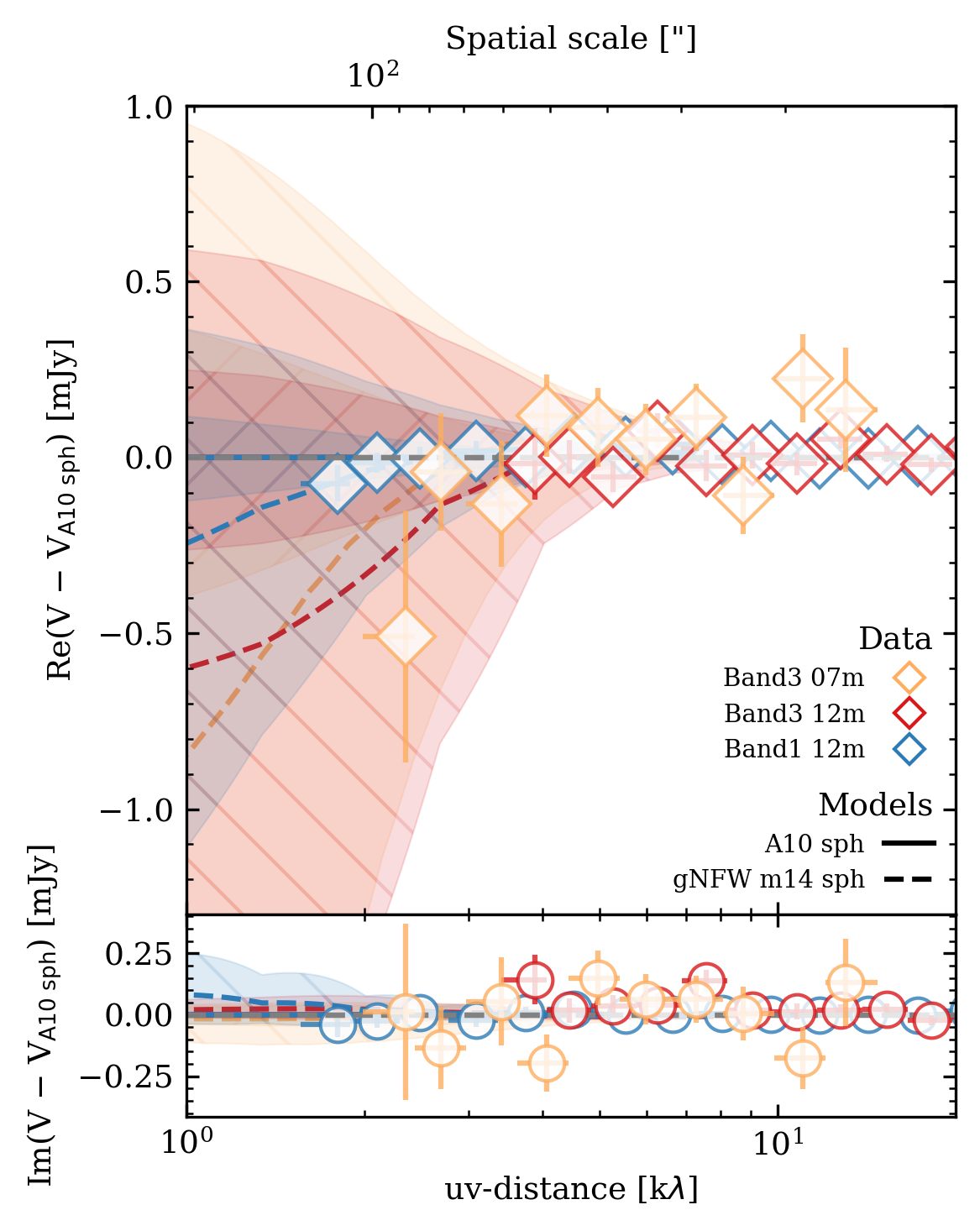}
                \caption{\textit{Left:} $uv$-radial distribution of the ALMA data and model, of which the visibilities are phase-shifted to the centre of the tSZ decrement (74.9230, -49.7818). The model is generated from taking a slice along $uv$-coordinates (0,$v$). The various colours represent the different ALMA configurations with respective transfer functions. To each of these datasets, we compare the respective forward model corresponding to the 50th percentile of the posterior distribution with a shaded area corresponding to the 16$^{\rm th}$-84$^{\rm th}$ percentile range. Thus, we highlight 2 models, the A10 spherical- (solid) and gNFW M14 sph (dashed) model (see Table~\ref{tab:params}), both corrected for the various transfer functions. \textit{Right:} Residual of the data, subtracted in the 2D $uv$-plane with the A10-spherical UPP model (top row Table~\ref{tab:params}). The residual flux shows that this high redshift cluster does not follow the ``Universal'' pressure profile.}
                \label{fig:uvradial}
            \end{figure*}
  
            \begin{figure}
                \centering
                \includegraphics[width=\linewidth]{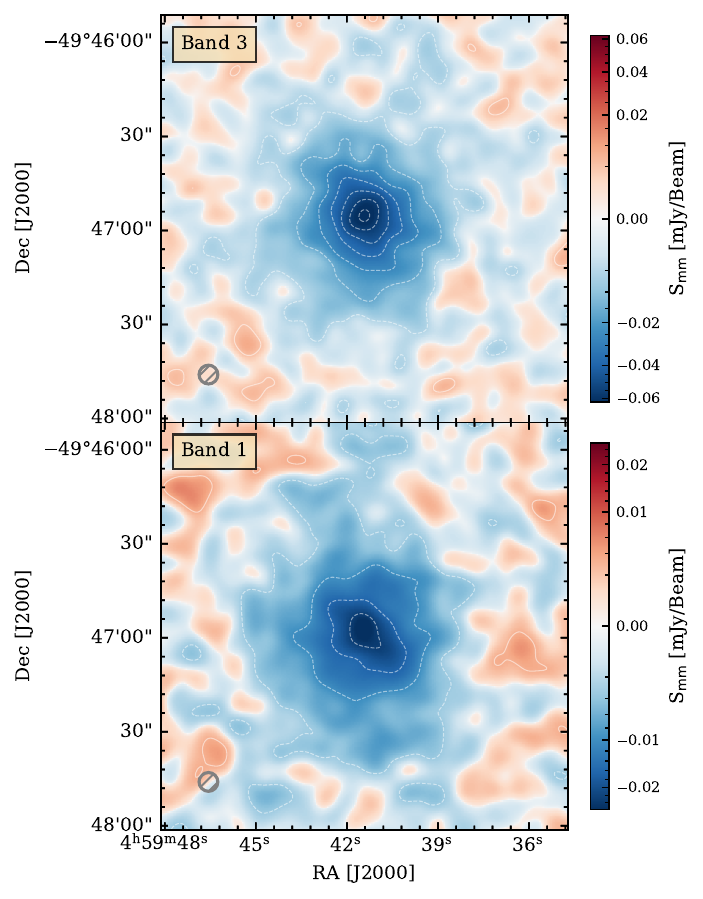}
                \caption{Continuum image reconstruction of CL~J0459 for Band 3 and Band 1. Contours start at $2\sigma$ and decrease after with a step size of $-4\sigma$. The peak SNR of both observations is $-27$ and $-18$, respectively, after applying a taper corresponding to a beam with FWHM of $6\arcsec$. The figures use a square-root scaling.}
                \label{fig:JvM_clean}
            \end{figure}
        
            The marginalised posterior parameters for both pressure models, assuming ellipsoidal and spherical morphologies, are listed in Table~\ref{tab:params}. For comparison, the ACT catalogue reports a halo mass of $M_{500,\mathrm{c}} = 3.4^{+0.5}_{-0.4} \times 10^{14}\,\mathrm{M}_\odot$ for CL~J0459 using the A10--UPP scaling \citep{Hilton2026}, while the SPT catalogue finds $M_{500,\mathrm{c}} = 3.1^{+0.5}_{-0.5} \times 10^{14}\,\mathrm{M}_\odot$ \citep{Bleem2024}. Our spherically symmetric maximum-posterior A10 model, marginalizing over ALMA calibration uncertainties, yields $M_{500,\mathrm{c}} = 2.2^{+0.3}_{-0.2} \times 10^{14}\,\mathrm{M}_\odot$. The quoted uncertainties are dominated by the intrinsic scatter in converting $\rm Y_{\rm 500,c}$ to $M_{\rm 500,c}$ following \citet{Arnaud2010} and \citet{Hilton2021}. The statistical precision is at the level of $\rm\approx 4\times10^{12}\,~M_\odot$, $\approx2\%$. While the A10 profile does not provide the best description of our data, we adopt it for the mass estimate to remain consistent with the ACT mass derivation, while benefiting from an improved handling of the contaminating compact sources. 

            We decided to omit halo masses when modelling the surface brightness with the gNFW formalism in Table~\ref{tab:params}. The tSZ effect directly probes the integrated thermal energy along the line of sight, and is not a direct probe of the total halo mass. Pushing tSZ observations to higher redshifts increases the uncertainty in the scaling relation used to convert $\rm Y_{500,c}$ to $\rm M_{500,c}$. Hence, we opted for one method with clear assumptions to estimate the tSZ-derived halo mass. The goal of this work is not to achieve the best tSZ-derived halo mass, but rather to perform a morphological analysis of the surface brightness distributions. In that regard, we found that a single-halo model provides a good description of the data, with no evidence requiring an additional halo component. 

            Based on the Bayesian evidence differences, we further conclude that the 2D distribution is well described by an ellipsoidal model at a significance of $>4\sigma$. Letting the normalisation and scale radius vary freely, rather than fixing them through the A10 formalism, is preferred at $\sim3\sigma$ pushing the scale radius to larger radii, while the recovered $\gamma$ remains consistent with the A10 value, albeit slightly higher. Beyond that, the data do not provide enough constraining power to warrant varying multiple shape parameters simultaneously: freeing $\beta$ and $\alpha$ in addition to $\gamma$ changes the evidence by less than $3\sigma$, and the recovered values remain consistent with the fixed ones.

            The forward model, evaluated in the native ALMA observing plane, is shown in Figure~\ref{fig:uvradial}. It displays the radially averaged visibilities of the pointing closest to the tSZ centroid, phase-shifted to that position for both the data and the model. Because the full two-dimensional $uv$ distribution is modelled directly, this figure is used for visualisation purposes only. Since the Band~3 observations consist of mosaics, we model the individual pointings separately and combine them atf the likelihood level. The right panel of Figure~\ref{fig:uvradial} shows the residual of the data, subtracted in the 2D $uv$-plane with the A10-spherical UPP model. 
            
            Finally, we use the A10-based model to deconvolve the ALMA data following \citet{vanMarrewijk2024}. The maximum-likelihood model, including contaminating point sources, is subtracted from the visibilities and the residuals are imaged. We then add the A10 model smoothed to a synthesised beam of $6\arcsec$, multiply by the primary beam, and combine with the imaged residual map. Since the data is forward modelled to both interferometric and single dish, the above described deconvolution functions as a data-combination method, akin to feathering, but without the need to first deconvolve the ACT Compton-$y$ map. The deconvolved, mosaicked continuum maps for each ALMA band are shown in Figure~\ref{fig:JvM_clean}, with peak S/N ratios of $-27$ and $-18$ in the Band~3 and Band~1 data, respectively. 
            
        \subsubsection{The radial density, temperature, and entropy profiles}\label{sec:fit_Chandra}
            
            We use \texttt{MBProj2D} \citep{Bulbul2024,Sanders2026} to forward-model the 2D X-ray surface brightness and infer radial electron density, temperature, and entropy profiles. Following \citet{Scheck2023}, we combine previously derived electron pressure $P_\text{e}$ profile measurements, in our case from the ALMA+ACT measurements, with the modelled electron number density $n_\text{e}$ to infer $T_\text{e}$ and $K_\text{e}$ as a function of radius directly. In contrast to \citet{Scheck2023}, we drop the hydrostatic equilibrium assumption and instead simply apply the ideal gas law,
            \begin{equation}\label{eq:gaslaw}
                P_\mathrm{e} = n_\mathrm{e}\, k_\textsc{b}\, T_\text{e},
            \end{equation}
            to infer the temperature and entropy profiles. Here, we adopt the pseudo-entropy definition of, 
            \begin{equation}
                 K_\text{e} = \frac{k_\textsc{b}T_\text{e}}{n_\text{e}^{2/3}}.
            \end{equation}
            
            The forward model is set up as follows: i) For the electron density profile $n_\text{e}$ we adopted the parametrised of \citet{Vikhlinin2006}, defined as,
            \begin{equation}
                \begin{split}
                    n_e(r) = n_{e,0}
                    \left(\frac{r}{r_c}\right)^{-\alpha/2}
                    \left(1 + \frac{r^2}{r_c^2}\right)^{-3\beta/2\,+\,\alpha/4} \\
                    \times\left(1 + \left(\frac{r}{r_s}\right)^{\gamma}\right)^{-\varepsilon/(2\gamma)},
                \end{split}
            \end{equation}
            where $n_{e,0}$ is the electron density normalisation, $r_c$ is the core radius, $\alpha$ modifies the inner cusp slope and is constrained to $\alpha \geq 0$ for the combined fit with X-ray and tSZ data (referred to as \textit{Chandra}+ALMA), $\beta$ governs the logarithmic slope of the outer profile, $\varepsilon$ controls the steepening beyond the scale radius $r_s$, and the transition sharpness $\gamma$ is fixed to $3$.
            ii) The cluster centroid is fixed to the tSZ centroid, which is found to be co-spatial within a pixel size of $2\arcsec$ with the X-ray centroid. 
            iii) We froze the X-ray ellipticity to that measured in the mm-wave regime and adopted the \texttt{eszee} line-of-sight geometry assumption (see Section~\ref{sec:fit_ALMA}) in \texttt{MBProj2D}.
            iv) The metallicity is fixed at $Z = 0.3\,\mathrm{Z}_\odot$ \citep{Mantz2020}. 
            v) The background is modelled as a flat component per energy band, normalised by the exposure map. 
            vi) Given \textit{Chandra}'s $\approx$half-arcsecond on-axis PSF and $\approx 2\arcsec$ bin size, we found the impact of PSF convolution to be negligible and is thus not included. 
            vii) Finally, we accounted for Galactic absorption by following \citet{Willingale2013}, taking into account both the molecular and atomic hydrogen components\footnote{https://www.swift.ac.uk/analysis/nhtot/index.php}, finding $n_\textnormal{H} = 1.23\times10^{20}~\textnormal{cm}^{-2}$ at the location of RA, Dec = $(74.91078, -49.77570)$. 
            
            \begin{table}
            \centering
            \caption{Priors on the parameters used for the X-ray fitting. Columns: (1) \textit{Chandra}, isothermal; (2) \textit{Chandra}, \citet{McDonald2014} $T(r)$; (3) \textit{Chandra}+ALMA.}
            \label{tab:priors_Xray}
            \footnotesize
            \setlength{\tabcolsep}{3pt}
            \begin{tabularx}{\columnwidth}{@{}l YYY@{}}
                \toprule
                Parameter & (1) & (2) & (3) \\
                \midrule
                \multicolumn{4}{@{}l}{\textit{Gas density, $n_{\rm e}(r)$}} \\
                $\ln(n_{\rm e,0}/{\rm cm^{-3}})$ & \multicolumn{3}{c@{}}{$\mathcal{U}(-14, 5)$} \\
                $\beta$ & \multicolumn{3}{c@{}}{$\mathcal{N}_{>0.2}(2/3, 0.3)$} \\
                $\ln(r_{\rm c}/{\rm kpc})$ & \multicolumn{3}{c@{}}{$\mathcal{N}(\ln 100, 1)$} \\
                $\alpha$ & \multicolumn{3}{c@{}}{$\mathcal{N}_{>0}(0, 1)$} \\
                $\epsilon$ & \multicolumn{3}{c@{}}{$\mathcal{N}(0, 1)$} \\
                $\ln(r_{\rm s}/{\rm kpc})$ & \multicolumn{3}{c@{}}{$\mathcal{N}(\ln 500, 0.5)$} \\
                $\gamma$ & \multicolumn{3}{c@{}}{3} \\
                \midrule
                \multicolumn{4}{@{}l}{\textit{Temperature, $T(r)$}} \\
                $\ln(T/{\rm keV})$ & $\mathcal{U}(\ln 0.06, \ln 60)$ & -- & -- \\
                $\ln(T_0/{\rm keV})$ & -- & $\mathcal{U}(\ln 0.5, \ln 40)$ & -- \\
                $T_{\rm min}/T_0$ & -- & 0.49 & -- \\
                $r_{\rm cool}/R_{500}$ & -- & 0.10 & -- \\
                $a_{\rm cool}$ & -- & 2 & -- \\
                $r_{\rm t}/R_{500}$ & -- & 0.38 & -- \\
                $a$ & -- & 0 & -- \\
                $b$ & -- & 3.26 & -- \\
                $c$ & -- & 0.94 & -- \\
                \midrule
                \multicolumn{4}{@{}l}{\textit{Cluster geometry}} \\
                RA [deg] & \multicolumn{2}{c}{74.92300} & 74.92259 \\
                Dec [deg] & \multicolumn{2}{c}{$-49.78190$} & $-49.78223$ \\
                $e$ & $\mathcal{U}(0.1, 1)$ & $\mathcal{U}(0.1, 1)$ & 0.826 \\
                $\theta$ [deg] & $\mathcal{U}(0, 360)$ & $\mathcal{U}(0, 360)$ & 75.9 \\
                \midrule
                \multicolumn{4}{@{}l}{\textit{Metal abundance}} \\
                $Z$ [Z$_\odot$] & \multicolumn{3}{c@{}}{0.3} \\
                \midrule
                \multicolumn{4}{@{}l}{\textit{Point sources}} \\
                $x,\,y$ [arcsec] & \multicolumn{3}{c@{}}{$x_0 \pm 3^{\prime\prime}$} \\
                $\ln\mathcal{A}$ & \multicolumn{3}{c@{}}{$\mathcal{U}(-25, 0)$} \\
                $\Gamma$ & \multicolumn{3}{c@{}}{1.7} \\
                \midrule
                \multicolumn{4}{@{}l}{\textit{Background}} \\
                $\ln b$ & \multicolumn{3}{c@{}}{$\mathcal{U}(-\infty, \infty)$} \\
                \bottomrule
            \end{tabularx}
            \begin{minipage}{\columnwidth}
                \vspace{2pt}
                \textbf{Notes.} $\mathcal{N}_{>a}(\mu,\sigma)$ is a Gaussian prior truncated to values above $a$. Parameters given as a single value are held fixed. Parameters sampled logarithmically are listed as $\ln(X/\mathrm{unit})$ and use natural logarithms. The gas density follows \citet{Vikhlinin2006}; the temperature shape parameters of column (2) are the $0.6<z<1.2$ values of \citet{McDonald2014}. Column (3) fixes the gNFW electron-pressure profile, centre, $e$ and $\theta$ to the ALMA SZ posterior medians (gNFW $\alpha\beta\gamma$-free, elliptical).
            \end{minipage}
        \end{table}
            
            Three compact X-ray sources, identified through visual inspection of the ICM-subtracted \textit{Chandra} image, are detected within the central $1\arcmin \times 1\arcmin$ region used for the 2D fit.  These are treated purely as contaminants of the extended cluster emission, independently of the mm-wave continuum and line sources discussed in Section~\ref{sec:cont_sources}. The X-ray sources are simultaneously modelled with the ICM-component as power-law emitters for which we model the amplitude and centroid. 
            The likelihood is evaluated pixel-by-pixel, and the posterior is sampled via MCMC, using the \texttt{emcee}-code \citep{emcee2013}. 

            As a baseline comparison, we also run the standard \texttt{MBProj2D} approach of inferring $T_\text{e}$ from broad, multi-band imaging, assuming a thermal bremsstrahlung model to model the amplitudes in the various energy bands jointly. We refer to this fit as the ``\textit{Chandra}''-fit. For this, we produce images in three broad energy bands (0.5--1.2, 1.2--2.5, and 2.5--6.0\,keV) using the calibrated X-ray events files. The electron pressure-constrained fit instead uses the single 0.5--2.0\,keV band. As for the temperature profile, we experiment with both the isothermal temperature profile model and the high redshift temperature template of \citet{McDonald2014}. All parameters with accompanied priors are summarized in Table~\ref{tab:priors_Xray}.
            
        \subsubsection{Interpretation of the modelled X-ray~\&~mm-wave observations}\label{sec:X-ray_interp}

            \begin{table}
                \centering
                \caption{Posterior parameters of the X-ray fits. Columns: (1) \textit{Chandra}, isothermal; (2) \textit{Chandra}, \citet{McDonald2014} $T(r)$; (3) \textit{Chandra}+ALMA.}
                \label{tab:params_Xray}
                \footnotesize
                \setlength{\tabcolsep}{3pt}
                \begin{tabular}{lccc}
                    \toprule
                    Parameter                            & (1)                    & (2)                    & (3)                    \\
                    \midrule
                    $n_{\rm e,0}$ [$10^{-2}$\,cm$^{-3}$] & $2.4^{+0.4}_{-0.4}$    & $2.5^{+0.4}_{-0.5}$    & $2.0^{+0.4}_{-0.4}$    \\
                    $\beta$                              & $0.69^{+0.11}_{-0.07}$ & $0.62^{+0.09}_{-0.06}$ & $0.66^{+0.13}_{-0.09}$ \\
                    $r_{\rm c}$ [kpc]                    & $101^{+22}_{-17}$      & $88^{+22}_{-16}$       & $112^{+32}_{-23}$      \\
                    $\alpha$                             & $0.24^{+0.25}_{-0.18}$ & $0.26^{+0.27}_{-0.19}$ & $0.21^{+0.25}_{-0.15}$ \\
                    $\epsilon$                           & $-1.5^{+0.9}_{-0.8}$   & $-1.2^{+0.8}_{-0.9}$   & $-0.4^{+1.0}_{-0.9}$   \\
                    $r_{\rm s}$ [kpc]                    & $450^{+180}_{-130}$    & $460^{+210}_{-140}$    & $480^{+310}_{-190}$    \\
                    $T$ [keV]                            & $10.2^{+2.1}_{-1.5}$   & --                     & --                     \\
                    $T_0$ [keV]                          & --                     & $14.2^{+3.3}_{-2.4}$   & --                     \\
                    $e$                                  & $0.67^{+0.05}_{-0.05}$ & $0.66^{+0.05}_{-0.04}$ & --                     \\
                    $\theta$ [deg]                       & $84^{+4}_{-4}$         & $82.9^{+3.9}_{-3.8}$   & --                     \\
                    \bottomrule
                \end{tabular}
                \begin{minipage}{\columnwidth}
                    \textbf{Notes.} Values are posterior medians with 16th/84th percentile uncertainties. ``--'' marks a parameter that is fixed in, or absent from, that model. Point-source and background parameters are omitted.
                \end{minipage}
            \end{table}

           \begin{figure*}
                \centering
                \includegraphics[width=\linewidth]{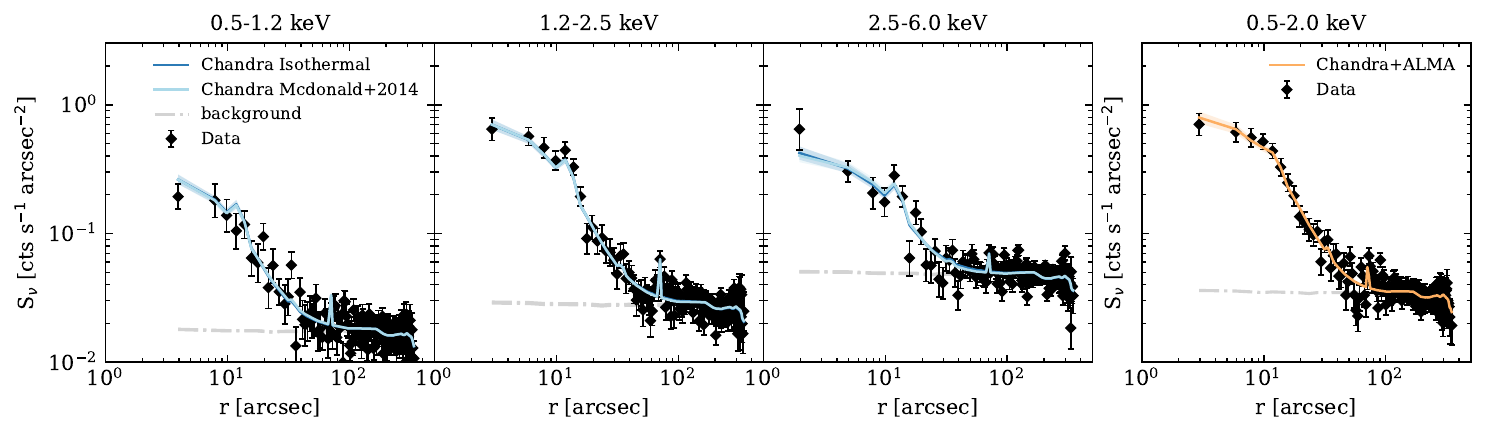}
                \caption{Radial X-ray surface brightness profiles of CL~J0459 in the three Chandra energy bands (0.5--1.2, 1.2--2.5, 2.5--6.0\,keV; \textit{left}) and the merged 0.5--2.0\,keV band (\textit{right}). Black diamonds show the data; coloured lines show the model median with uncertainties based on the $16^\text{th}-84^\text{th}$ percentile range; the dashed-dotted gray line marks the modelled background. Note that the likelihood is evaluated on a pixel-by-pixel basis in the image, not on these radially binned profiles; the binning (equal to the pixel size) of both data and projected model is performed only in post-processing for visualisation. The non-smooth features seen here arise from this binning and from the compact sources, which are included in both the data and model.}                
                \label{fig:Xrayprofiles}
            \end{figure*}
            
            \begin{figure*}
                \centering
                \includegraphics[width=\linewidth]{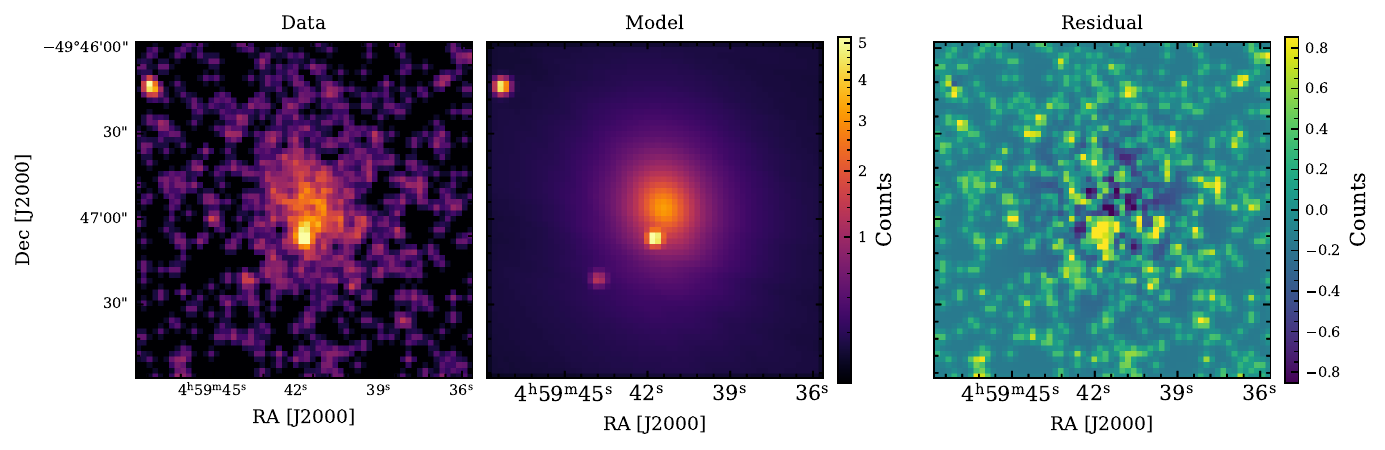}
                \caption{Smoothed \textit{Chandra} X-ray surface brightness maps of CL~J0459 in the merged 0.5--2.0\,keV band: obtained data (left), most-likely 2D elliptic Vikhlinin model (center), and residuals (right). A slight flux excess is visible to the south of the cluster core, which is off-centred from the bulk of the ICM.}
                \label{fig:2Dfitting}
            \end{figure*}
    
            \begin{figure*}
                \centering
                \includegraphics[width = \hsize]{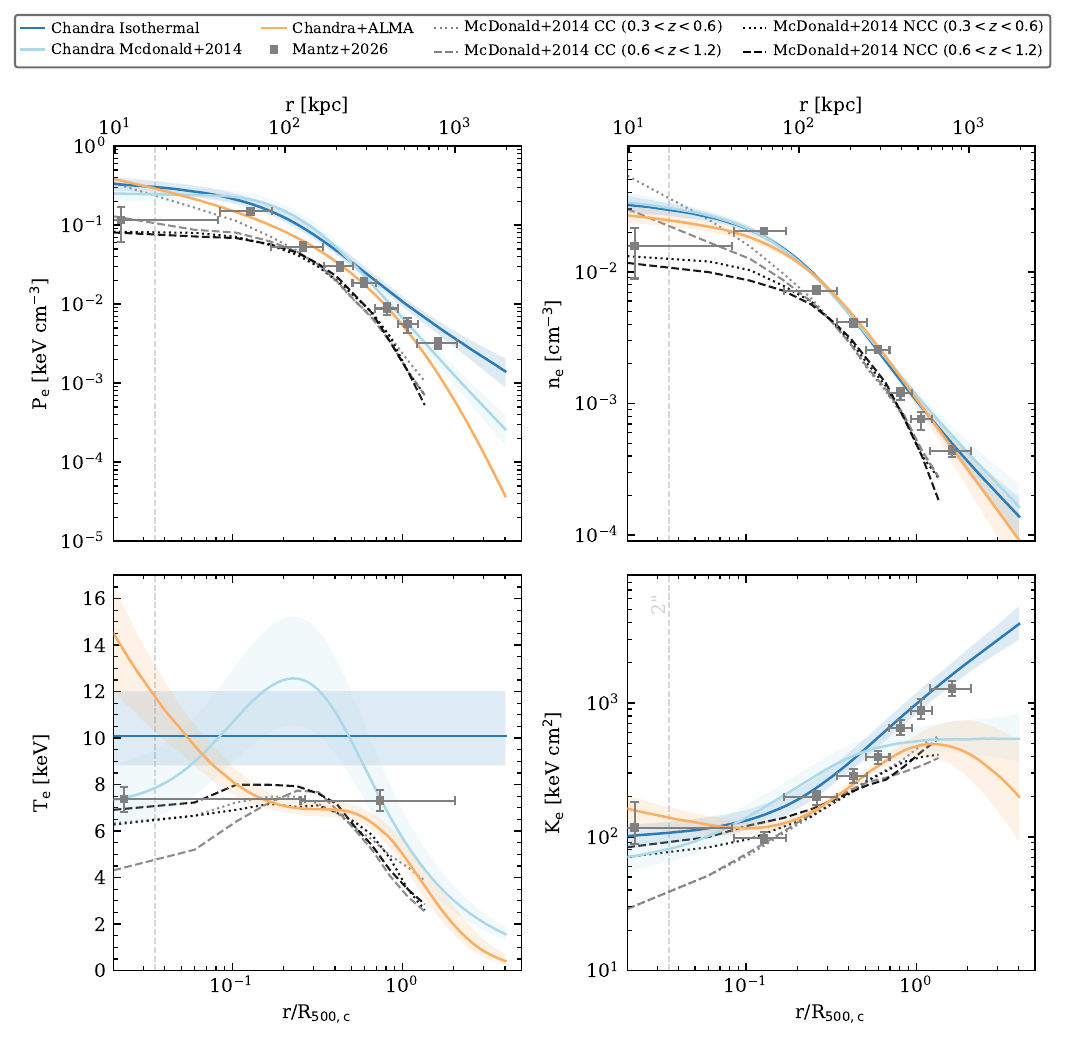}
                \caption{Radial thermodynamic profiles of CL~J0459: electron pressure $P_\text{e}$, number density $n_\text{e}$, temperature $T_\text{e}$, and entropy $K_\text{e}$. Results from broadband X-ray temperature fitting (blue: split in a model for which we assume an isothermal assumption and the temperature profile template from \citealt{McDonald2014}), and from combining \textit{Chandra} with ALMA+ACT (orange) are compared with \citet[][grey squares]{Mantz2026}. For comparison we added the population average profiles of \citet{McDonald2014} split in morphological state (cool-cores, CC, light-gray; non-cool-cores, NCC, dark-gray; and total average, All, gray) and redshift (lower $z$, dotted lines; intermediate $z$, dash-dotted). The vertical dashed line marks minimum recoverable scale from ALMA (Fig.~\ref{fig:sensitivity}).}
                \label{fig:phys_profiles}
            \end{figure*}

            The marginalised posterior parameters for modelling done on the X-ray data are listed in Table~\ref{tab:params_Xray}. The inferred surface brightness profiles for both methods are shown in Figure~\ref{fig:Xrayprofiles}. The 2D \textit{Chandra}+ALMA model displayed next to the data and residuals are shown in Figure~\ref{fig:2Dfitting}; the residuals reveal a slight flux excess to the south of the cluster core, which is discussed further in Section~\ref{sec:sloshing}. To test whether the residual pattern shown in Figure~\ref{fig:2Dfitting} is an artefact of the model constraints, we performed a Chandra-only fit with all geometric parameters free (centroid, ellipticity, position angle); the southern excess and the negative northern residuals persist, so they are not driven by a forced centring or symmetry. The negative features in the residuals are probably an artifact of the low photon counts at this redshift combined with the smooth ICM model — one inevitably over- or under-subtracts in individual pixels. 
            
            The inferred thermodynamic profiles are shown in Figure~\ref{fig:phys_profiles} alongside the \textit{XMM-Newton}-based measurements of \citet{Mantz2026}, providing an independent cross-check; universal thermodynamic profiles from \citet{McDonald2014}, scaled to our inferred halo mass, are plotted for comparison. Note that the pressure profile of the \textit{Chandra}+ALMA run shown in the top-left panel is set to the maximum a posteriori model from the ALMA modelling and is thus constrained by ALMA+ACT analysis only. Here, we adopted the gNFW $\alpha\beta\gamma$-free (ell) model. Choosing another model, such as to the GNFW A10 $\gamma$-free model that was modelled with more restriction and only a minor Bayesian evidence increase of +3 with respect to the GNFW A10 $\gamma$-free model, resulted in no significant change on the thermodynamical profiles. 
            
            We find excellent agreement between our density profile and the \textit{XMM-Newton}-derived data points of \citet{Mantz2026}. The only disagreement is present in the core of the cluster, which stems from our physically motivated prior disallowing negative inner slopes in our parametric fit. The temperature profiles are less consistent: our inferred profile, obtained by freezing the electron pressure distribution to our ALMA+ACT model, shows a hot core that declines outward and plateaus over $r \sim 0.1$--$0.7\,r_{500,c}$, and it remains discrepant with the isothermal estimate of \citet{Mantz2026} and the universal temperature profile of \citet{McDonald2014} -- both the assembled average values (shown in gray) and our own \textit{Chandra}-only inference (light blue), for which we adopt the \citet{McDonald2014} model for the NCC, high redshift profile. For the isothermal \textit{Chandra}-only run (dark blue), we find a higher overall temperature, $k_{\rm B}T_{\rm e}=10\pm2$~keV, than that measured by \textit{XMM-Newton}, $k_{\rm B}T_{\rm e}=7.2\pm0.3$~keV \citep{Mantz2026}. We find this not to be an artifact of the isothermal assumption, given that our inferred radial temperature profile when assuming a \citet{McDonald2014} distribution also displays elevated temperatures over the radial range $0.1<r/r_{500,c}<1$. The resulting temperature ratio of $1.4 \pm 0.3$ between the isothermal \textit{Chandra}-only run and the \citet{Mantz2026} average temperature is consistent with the known calibration differences between \textit{Chandra} and \textit{XMM-Newton} \citep{Schellenberger2015, Migkas2024}. The interpretation of the temperature and entropy distributions is continued in Section~\ref{sec:profiles_discussion}.
            
    \subsection{Thermodynamic maps}\label{sec:maps}
    
        To investigate any asymmetric and off-centred features, we construct projected maps of the ICM thermodynamic quantities following the methodology of \citet{Adam2017} and \citet{Churazov2021}, with minor adjustments with respect to these original works. These maps are constructed by combining the Compton-$y$ map derived from the deconvolved ALMA data (which by construction are combined with the ACT data, see Section~\ref{sec:fit_ALMA}) with the \textit{Chandra} and XMM-Newton X-ray surface brightness maps, yielding spatially resolved distributions of thermal electron pressure, density, temperature, and entropy at $6\arcsec$ resolution. For both the ALMA and \textit{Chandra} data, point source contaminants are subtracted out of the maps based on the surface brightness modelling described in Section~\ref{sec:profiles}. The methodology of how the maps are constructed is outlined in the following subsections. 

        \subsubsection{Thermal pressure mapping}\label{sec:method_maps}
        
            The observable we exploit to map the thermal pressure of the ICM is the tSZ effect, as also mentioned in Section~\ref{sec:fit_ALMA}. The amplitude of the tSZ signal is given by
            \begin{equation}\label{eq:spectrum}
                \Delta I_{\nu} = I_0\,g(x, T_\text{e})\, y_\textsc{sz},
            \end{equation}
            where $I_0 \approx 270.33\,\left(T_{\textsc{cmb}}/2.7255\,\text{K}\right)^3 \, \text{MJy\,sr}^{-1}$ \citep{Fixsen2009}, $x = h\nu/(k_\textsc{b} T_{\textsc{cmb}}) \approx \nu/56.8~\rm GHz$, and $g(x, T_\text{e})$ is the tSZ spectral function. The Compton-$y$ parameter is defined as
            \begin{equation}\label{eq:comptony}
                y_\textsc{sz} = \frac{\sigma_\textsc{t}}{m_\mathrm{e}c^2} \int P_\mathrm{e}\, \text{d}l \equiv \frac{k_{\rm B}\,\bar{T}_\text{e}}{m_\mathrm{e} c^2}\,\sigma_\textsc{t} \int n_\mathrm{e}\, \text{d}l,
            \end{equation}
            where we use the ideal gas law, Eq.~\eqref{eq:gaslaw}, to link the electron pressure $P_\text{e}$ to the electron number density $n_\text{e}$ and temperature $T_\text{e}$, and $\sigma_\textsc{t}$, $m_\mathrm{e}$, $c$, and $l$ are the Thomson cross section, electron mass, speed of light, and line-of-sight path length, respectively. To bring $T_\text{e}$ outside the integral, we introduce the electron density-weighted electron temperature
            \begin{equation}\label{eq:Temw}
                \bar{T}_\text{e} = \frac{\int T_\text{e}\, n_\text{e}\, \text{d}l}{\int n_\text{e}\, \text{d}l}.
            \end{equation}
            Combining this with the electron density-weighted electron density,
            \begin{equation}\label{eq:nemw}
               \Bar{n}_\text{e} = \frac{\int n_\text{e}^2 \text{d}l}{\int n_\text{e}\, \text{d}l},
            \end{equation}
            the measured Compton-$y$ can be converted to an effective line-of-sight pressure via
            \begin{equation}\label{eq:P_int}
                \Bar{P}_\text{e} = \Bar{n}_\text{e}\, k_\textsc{b}\, \bar{T}_\text{e} = \frac{\int n^2_\text{e}\, \text{d}l}{\left(\int n_\text{e}\, \text{d}l\right)^2}\int P_\text{e}\, \text{d}l \equiv \frac{1}{l_\text{eff}}\int P_\text{e}\, \text{d}l,
            \end{equation}
            and equivalently,
            \begin{equation}\label{eq:P_map}
                \Bar{P}_\text{e} = \frac{m_\text{e}c^2}{\sigma_\textsc{t}}\, \frac{y_\textsc{sz}}{l_\text{eff}},
            \end{equation}
            where, following \citet{Adam2017} and \citet{Wicker2026}, the effective path length is defined as
            \begin{equation}\label{eq:leff}
                l_\text{eff} \equiv \frac{\left(\int n_\text{e}\text{d}l\right)^2}{\int n^2_\text{e}\text{d}l},
            \end{equation}
            \noindent which probes the length of the cluster along the line of sight. Note that $l_\text{eff}$ is a direct consequent of moving the thermodynamic quantities, such as $P_\text{e}$, outside of the line of sight integral. Hence, this quantity does not enter into the computation of the thermodynamic radial profiles previously discussed in Section~\ref{sec:profiles}.
            
        \subsubsection{Electron density mapping}
        
            The average electron density along the line of sight can also be expressed in terms of $l_\text{eff}$, following Eqs.~\eqref{eq:nemw} and \eqref{eq:leff}, as
            \begin{equation}\label{eq:n_eff}
                \Bar{n}_\text{e} = \frac{1}{l_\text{eff}} \int n_\text{e}\,\text{d}l.
            \end{equation}
            Assuming that the X-ray emission is dominated by thermal bremsstrahlung, the X-ray surface brightness can be written as
            \begin{equation}\label{eq:X-ray_brightness}
                \begin{aligned}
                    S_\text{X} &= \frac{1}{4\pi(1+z)^4\cdot\mathrm{ECF}}\int n^2_\text{e}\,\Lambda\left(T_e,Z_e\right)\,\text{d}l \\
                               &\approx \frac{\Lambda\left(\left<T_e,Z_e\right>\right)}{4\pi(1+z)^4\cdot\mathrm{ECF}}\int n^2_\text{e}\,\text{d}l,
                \end{aligned}
            \end{equation}
            where $z$ is the cluster redshift and $\Lambda(T_e, Z_e)$ is the cooling function in the relevant energy band.
            The derived maps are not the product of an approach that facilitates modelling the temperature distribution along each line of sight. Hence, unlike our profile forward modelling (Section \ref{sec:profiles}), we treat  $\Lambda(T_e, Z_e)$ as constant --- a good approximation for dense clusters \citep{Churazov2021} --- for the map-based quantities reported in this section. Finally, $\text{ECF}$ is the energy conversion factor used to convert count rates to fluxes. Combining Eqs.~\eqref{eq:n_eff} and \eqref{eq:X-ray_brightness} then yields
            \begin{equation}\label{eq:neff}
                \Bar{n}_\text{e} = \sqrt{\frac{4\pi(1+z)^4\,F_\text{X}}{l_\text{eff}\,\Lambda\left(\left<T_e,Z_e\right>\right)}},
            \end{equation}
            which expresses the effective electron density solely in terms of the X-ray surface brightness, analogous to Eq.~\eqref{eq:P_map} for the thermal pressure. Here, we defined $F_\text{X}=S_\text{X}\cdot\text{ECF}$.

            To convert the \textit{Chandra} and XMM-Newton count rate measurements to fluxes, we use \texttt{PyXspec} \citep{Arnaud1996, Gordon2021} together with the response files generated for each observation over the source extraction region, the \textsc{rmf} and \textsc{arf} files, to compute the ECF. If the \textsc{rmf} and \textsc{arf} files are not available we reverted to using \texttt{webPIMMS}\footnote{\url{https://heasarc.gsfc.nasa.gov/cgi-bin/Tools/w3pimms/w3pimms.pl}} assuming an APEC (Astrophysical Plasma Emission Code; \citealt{Smith2001}) model with $T_e = 7.2$~keV, $Z = 0.4\,\mathrm{Z}_\odot$, and $z = 1.71$, the Galactic absorption of $n_\textnormal{H} = 1.23\times10^{20}~\textnormal{cm}^{-2}$ at RA, Dec $= (74.91078, -49.77570)$, and compute it considering the energy rate of $0.5-2.0$\,keV for \textit{Chandra} and $0.4-4.0$\, keV for XMM-Newton -- matching the full energy band considered of the specific observation. 
            The ECF values used in this analysis are:
            \[
            \mathrm{ECF}_{\mathrm{Chandra}} = 9.83 \times 10^{-12}\,\mathrm{erg\,cm^{-2}\,ct^{-1}},
            \]
            \[
            \mathrm{ECF}_{\mathrm{XMM-Newton}} = 5.572 \times 10^{-12}\,\mathrm{erg\,cm^{-2}\,ct^{-1}},
            \]
            for which we used the \texttt{PyXspec}-derived ECF for the \textit{Chandra} data and the \texttt{webPIMMS} one for XMM-Newton. 
            
            To compute $\Lambda\left(\left<T_e,Z_e\right>\right)$, we used the \texttt{pyatomdb} code \citep{Foster2012}, assuming a collisional ionisation equilibrium plasma. Metal abundances (elements with atomic number $Z \geq 3$) are set to $0.37\,\mathrm{Z}_\odot$. For each observed-frame energy bin $[E_1, E_2]$, we query \texttt{pyatomdb} in the corresponding rest-frame band $[E_1(1+z), E_2(1+z)]$, integrate the emitted photon spectrum over the bin, and convert to units of erg~cm$^3$~s$^{-1}$ using the bin-centre photon energy. The cooling function is tabulated over 1001 temperatures log-spaced between $10^5$ and $10^9$~K and interpolated at the spectroscopic temperature $\langle T_\text{e} \rangle$ during the map computation. The average X-ray background flux is subtracted from all input maps prior to further analysis.

            \begin{figure*}
                \centering
                \includegraphics[width=0.92\textwidth]{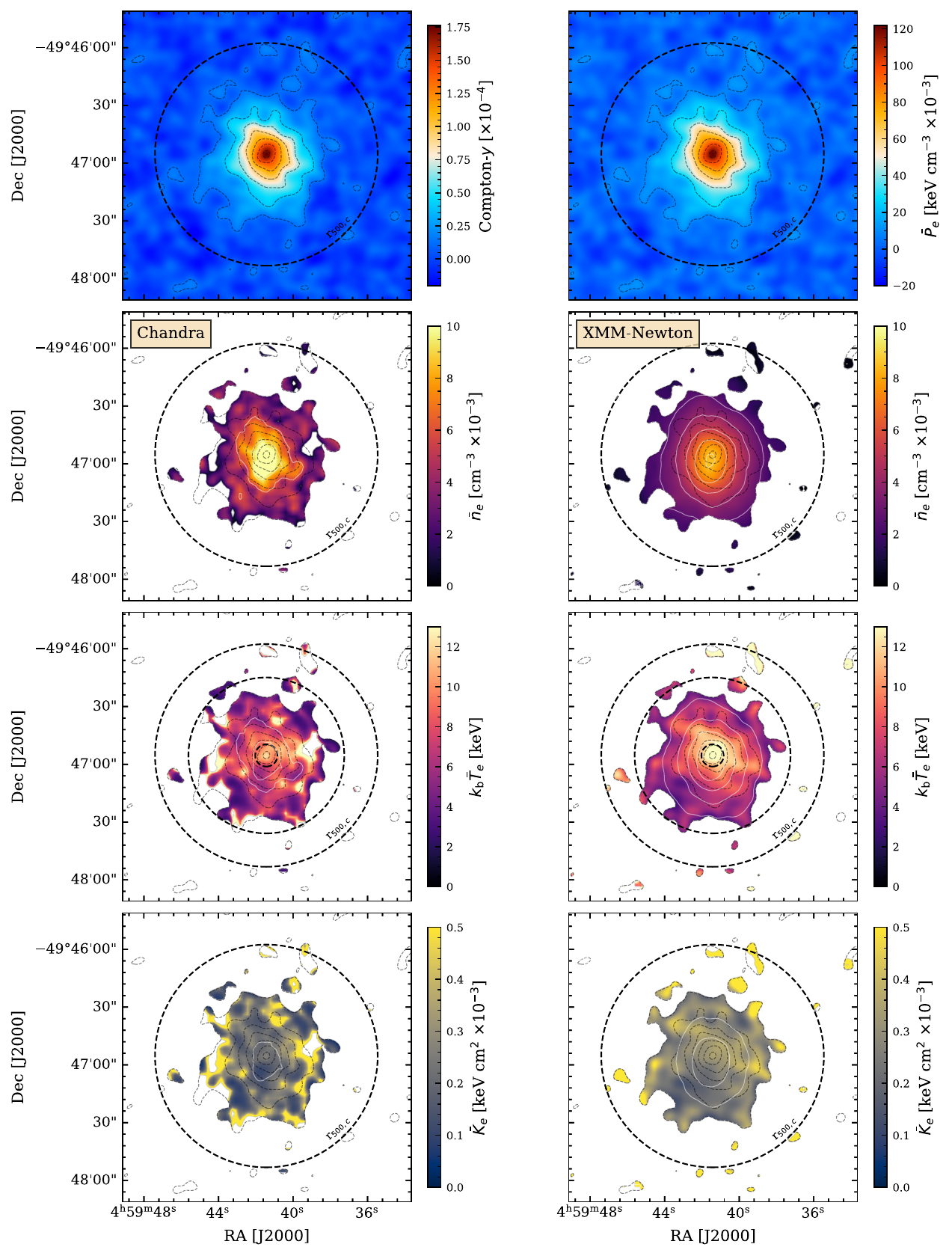}
                \caption{Projected thermodynamic properties of CL~J0459 derived from ALMA+ACT tSZ measurements and \textit{Chandra} (left) and XMM-Newton (right) data. See Section~\ref{sec:dynamics} for how the quantities are computed. Pixels with Compton$-y$ S/N~$< 2\,\sigma$ are masked. Black contours from Fig.~\ref{fig:JvM_clean} are overlaid; white contours show the per-quantity S/N at $[1.5,2.5,3.5,4.5,5.5]\,\sigma$. We further annotate $\rm r_{500,c}$ with the dashed circles. For the temperature maps, we also show $\rm r=0.1,\,0.7\,r_{500,c}$, to highlight the region where the temperature profiles in Figure~\ref{fig:phys_profiles} plateau.}
                \label{fig:maps}
            \end{figure*}
            
            \subsubsection{Temperature \& Entropy Mapping}\label{sec:temp_theory}
                
                Taking the ratio of Eqs.~\eqref{eq:P_map} and \eqref{eq:neff} and applying the ideal gas law, Eq.~\eqref{eq:gaslaw}, yields a density-weighted temperature map,
                \begin{equation}\label{eq:teff}
                    k_\textsc{b}\bar{T}_\text{e} = \frac{\Bar{P}_\text{e}}{\Bar{n_\text{e}}} = \frac{y_\textsc{sz}}{\sqrt{F_\text{X}}}\frac{1}{\sqrt{l_\text{eff}}}\frac{m_\text{e}c^2}{\sigma_\text{T}}\sqrt{\frac{\Lambda\left(\left<T_\text{e},Z_\text{e}\right>\right)}{4\pi(1+z)^4}}.
                \end{equation}
                Similarly, we define a pseudo-entropy parameter as
                \begin{equation}\label{eq:k_eff}
                    \Bar{K} = \frac{k_\textsc{b}\bar{T}_\text{e}}{\Bar{n}_\text{e}^{2/3}} = \frac{m_\text{e}c^2}{\sigma_\textsc{t}}\,\left(\frac{4\pi(1+z)^4\,F_\text{X}}{\Lambda\left(\left<T_e,Z_e\right>\right)}\right)^{5/6} \frac{y_\textsc{sz}}{l^{11/6}_\text{eff}},
                \end{equation}
                which is sensitive to contrasts in the ICM thermodynamic structure and serves as a useful diagnostic for the presence of turbulence or mixing.

            \subsubsection{Parametrisation of $l_{\text{eff}}$}\label{sec:leff_method}
            
                The largest systematic uncertainty in Eqs.~\eqref{eq:teff} and \eqref{eq:k_eff} arises from the parametrisation of $l_\text{eff}$, for which we explore two approaches. The first assumes a constant value, normalised such that the median temperature within $30\arcsec$ equals the spectroscopic isothermal temperature of \citet{Mantz2026}, $k_\text{b}T_\text{e}=7.2$\,keV. This is equivalent to normalising such that the integral within an arbitrary radii of the X-ray pseudo-pressure map equals the integrated Compton-Y as often done in the literature \citep[see][]{Mroczkowski2012,Sayers2013}. We found $l_\text{eff}=360\pm30$\,kpc. Here, we propagated the uncertainty on the spectroscopic temperature fit of $\pm0.3$\,keV through the $l_\text{eff}$ estimate using Monte-Carlo sampling.

                Our second approach follows \citet{Churazov2021} and adopts the best-fitting $\beta$-profile for the radial electron density from \citet{Mantz2020} to construct a projected $l_\text{eff}$ map via Eq.~\eqref{eq:leff}. The full derivation is given in Appendix~\ref{app:leff}, with the result
                \begin{equation}\label{eq:l_eff_projected}
                    l_\text{eff} = \frac{r_\text{c}\sqrt{\pi}}{2}\left(1+\frac{R^2}{r_\text{c}^2}\right)^{1/2}\frac{\Gamma\left(3\beta/2-1/2\right)^2\Gamma(3\beta)}{\Gamma(3\beta/2)^2\Gamma(3\beta-1/2)},
                \end{equation}
                with $\beta > 1/3$. Notably, $\beta$ enters only as an amplitude parameter, leaving $r_\text{c}$ as the primary determinant of the spatial variation of $l_\text{eff}$: the distribution is flat in the inner region out to $r_\text{c}$, beyond which it declines linearly. For $r_\text{c} = 20\arcsec$ and $\beta = 0.82$, we find $l_\text{eff} = 364\,$kpc within the inner $20\arcsec$ (thus consistent with the data-driven method). Both parametrisations and their impact on the retrieved thermodynamic distributions are discussed in Section~\ref{sec:clump}.

        \subsubsection{Interpretation of Thermodynamic maps}\label{sec:projected_maps}
            
            The projected pressure, density, temperature, and pseudo-entropy maps are displayed in Figure~\ref{fig:maps} at a resolution of $6\arcsec$ (52\,kpc), constructed by combining the ALMA+ACT Compton-$y$ map with the \textit{Chandra} (left) and XMM-Newton (right) X-ray surface brightness maps. Prior to map construction, continuum point sources are removed from both the ALMA and \textit{Chandra} data based on their fitted positions and amplitudes from the surface brightness modelling described in Sections~\ref{sec:fit_ALMA} and \ref{sec:fit_Chandra}, respectively. For the map reconstruction, we use only the 90~GHz Band~3 Compton-$y$ map, as it has a S/N approximately twice as high as the Band~1 data and is less affected by continuum point source contamination, while yielding consistent fitted pressure profiles with the Band~1 data. 
            For all maps we adopt a constant value for $l_{\text{eff}}$.


            We further introduce a scaling factor \textsc{C}$_{\text{T}}$ to rescale the median projected temperature within $30\arcsec$ of the \textit{Chandra}-derived $k_\text{b}\bar{T}_{\rm e}$ map to its XMM-Newton counterpart, such that $k_{\rm b}\bar{T}_{\text{XMM},\text{e}} = \textsc{C}_{\text{T}}k_{\rm b}\bar{T}_{\textit{Chandra},\text{e}}$. We find \textsc{C}$_{\text{T}}=0.84$. Differences between the two instruments are expected, given the difference in angular resolution, with \textit{Chandra} being more sensitive to compact emission while less susceptible to unresolved gas clumping, as well as the difference in calibration as mentioned in section~\ref{sec:X-ray_interp}. We note that by following this procedure, we bias the amplitude of the 2D thermodynamic distributions to the spectroscopic temperature measurements of \citet{Mantz2026}. Future work will perform a full joint multi-wavelength fit across all instruments, including spectroscopic constraints, to properly account for geometry, clumpiness, and other systematic biases.
            
\section{Astrophysical Implications}\label{sec:discussion}

    CL~J0459 is a disturbed cluster. It has a dynamic ICM with a hot core that radially declines to lower temperatures. Although the temperature falls radially, the entropy distribution remains rather flat around $200$~keV\,cm$^{2}$ up to $r_{500,c}$. This nearly flat distribution with a short effective path length suggest that CL~J0459 is dynamically active, possibly thermalising its ICM. The next subsections will go into more detail on the interpretation of the measurements we have provided. In particular, we discuss the clumpiness of the ICM (Section~\ref{sec:clump}), possible presence of sloshing seen in the entropy distribution (Section~\ref{sec:sloshing}), impacts of the temperature and entropy profile (Section~\ref{sec:profiles_discussion}), the hydrostatic mass bias (Section~\ref{sec:mass_bias}), and how CL~J0459 integrated quantities compare with local observations via scaling relationships (Section~\ref{sec:scaling}).

    \subsection{A clumpy ICM}\label{sec:clump}
    
        To investigate whether gas clumping may contribute to the observed discrepancy between the \textit{Chandra} and XMM-Newton temperature estimates, we rewrite Eq.~\eqref{eq:leff} in terms of the clumpiness factor \citep{Birkinshaw1991, Eckert2015}, following \citet{Adam2017},
        \begin{equation}
            l_\text{eff} = \frac{r_{500,c}}{C^2\left(n_\text{e}\right)},
        \end{equation}
        where,
        \begin{equation}\label{eq:C}
            C^2\left(n_\text{e}\right) = \frac{\left<n^2_\text{e}\right>}{\left<n_\text{e}\right>^2}
        \end{equation}
        is the clumpiness parameter, when setting $\int~\text{d}l=r_\text{500,c}$ . Using our fitted $l_\text{eff}=360\pm30$\,kpc and $r_{500,c} = 490$~kpc, we obtain $C(n_\text{e}) = 1.17 \pm 0.05$, within a projected $\approx 0.5\,r_{500,c}$ since the \textit{Chandra} and XMM-Newton data do not probe further. This value is consistent with, though at the upper edge of, what is found in local clusters, where both simulations and observations find $C \lesssim 1.1$ within $r_{500,c}$ \citep{Vazza2013, Eckert2015, Planelles2016}, with higher values only reached in the cluster outskirts near $R_{200}$ \citep{Vazza2013}. A clumpiness at this level would be consistent with CL~J0459 being a dynamically active system at high redshift, where ongoing accretion and possible merging activity can drive gas inhomogeneities \citep{Planelles2016, Pratt2019}.    
        
    \subsection{Sloshing}\label{sec:sloshing}
    
        \begin{figure}
            \centering
            \includegraphics[width=0.98\linewidth]{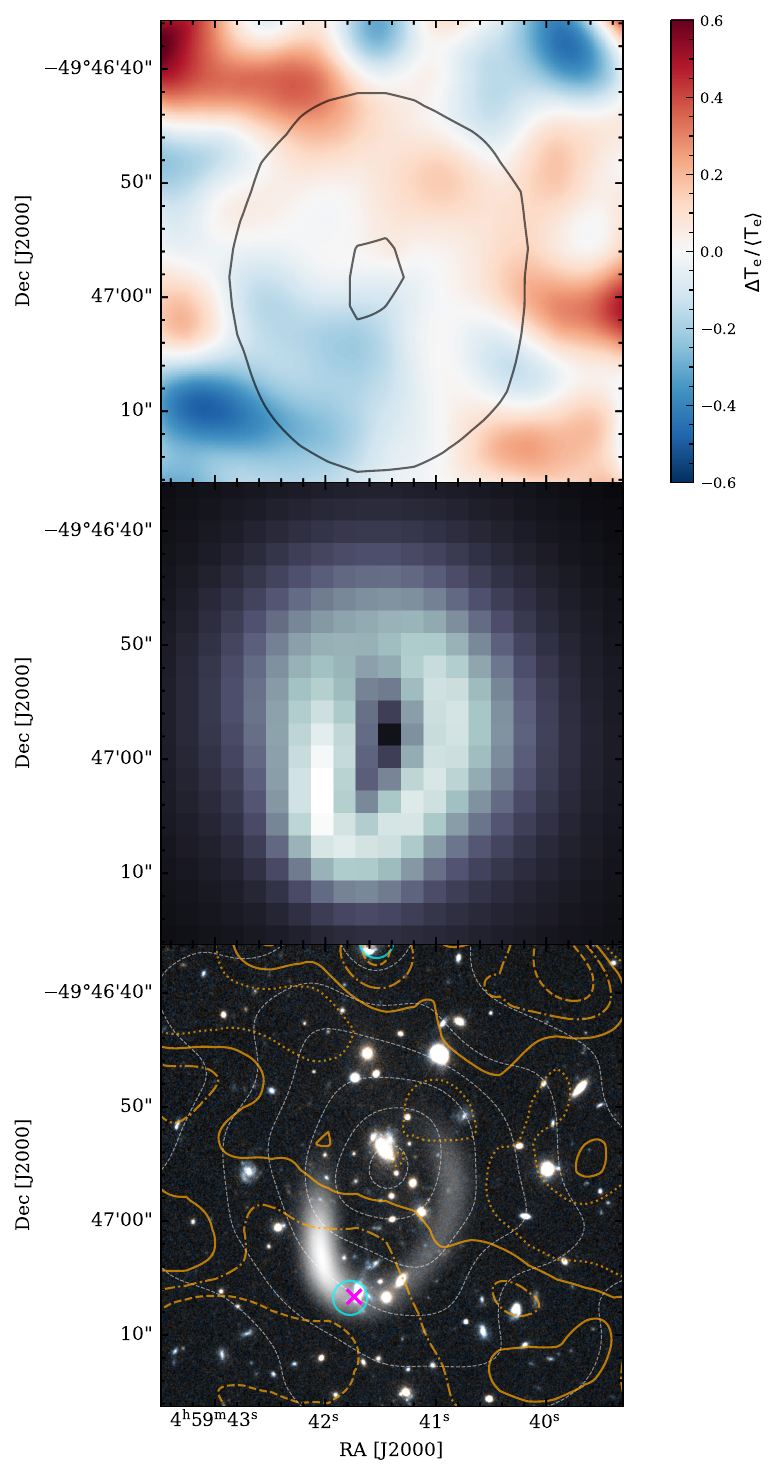}
            \caption{\textit{Top:} Fractional temperature difference, $\Delta T_e / \left<T_e\right>$, from XMM-Newton where we subtracted of the azimuthally-averaged radial profile. Black contours show the electron density significance at 3.5 and 5.5$\sigma$. \textit{Middle:} Adaptively smoothed \textit{Chandra} GGM image of CL~J0459. \textit{Bottom:} Zoomed-in HST background (F140W/F110W) of the CL~J0459 field, overlaid with electron entropy $K_e$ contours (XMM-Newton; orange) at 200, 250, 300, and 350\,keV\,cm$^2$ (dashed, dash-dotted, solid, and dotted, respectively); the GGM map from the middle panel (white); and the marked location of the CO(2--1) line (circle) and X-ray-detected AGN (cross). 
            The co-spatiality of the entropy dip, temperature gradient, disturbed galaxy population, and extended GGM detection is indicative of sloshing.}
            \label{fig:wedge}
        \end{figure}

        Using maps rather than solely focusing on radial surface brightness profiles opens up the study of morphological asymmetries and surface brightness discontinuities. In the thermodynamic distributions of CL~J0459, we do not observe any strong discontinuities that would hint at the presence of supersonic shock fronts. However, we do observe a dip in the electron temperature and entropy distribution to the south of the X-ray peak (cf. Figure~\ref{fig:maps}). This feature is most prominent in the XMM-Newton entropy map. 
        
        Figure~\ref{fig:wedge} highlights the cold spot in three ways: a residual map, obtained by subtracting the azimuthally averaged radial profile, which exposes deviations from spherical symmetry; an adaptively smoothed Gaussian Gradient Magnitude (GGM; \citealt{Sanders2016}) map computed from the \textit{Chandra} $0.5$--$2.0$~keV surface brightness map; and the bottom panel overlays the latter on the HST image from Figure~\ref{fig:sources}, together with contours of the XMM-Newton projected electron entropy map from Figure~\ref{fig:maps}.
        The entropy dip coincides with the residual flux in Figure~\ref{fig:2Dfitting}, indicating a locally enhanced density, while the lack of a counterpart in the Compton-$y$ map implies pressure equilibrium. Given that the single-halo model is elongated along the north-east to south-west axis, the southern residual flux may trace an on-axis merger seen as a sloshing, appearing as a gradient in the azimuthally averaged residual map; similar features are found in the local Universe and likewise attributed to sloshing \citep[see, e.g.,][]{Okabe2021}. An X-ray AGN co-spatial with a CO~J=2-1 line emitter is detected at this position and is likely to be an off-centred BCG in the merger remnant. At the same location the near-IR galaxy population is visibly disturbed, resembling the remnant core of a group colliding with the host.
        
        The presence of sloshing was previously suggested by \citet{McDonald2017}, based on the bright off-centred flux seen in the surface brightness. While the low significance of the entropy dip means this interpretation remains speculative, the inclusion of ALMA data consistent with pressure equilibrium, combined with resolved measurements of the temperature and entropy distributions, strengthens the case. Similarly, a sloshing cold front has been posited in SpARCS104922.6+564032.5 at $z=1.7$ \citep{HLavacek-Larrondo2025} which was invoked to explain the origin of a tentative mini radio halo. Resolved tSZ observations, carried out in a similar fashion to those presented in this work, could test this hypothesis. Vice versa, L-band MeerKAT data of CL~J0459 do not yet reveal a radio halo (M.\ Pannella et al., in prep.); deeper radio measurements or more thorough analysis would therefore be valuable for further studying the co-evolution of the thermal and non-thermal ICM.
        
    \subsection{Entropy and temperature distributions}\label{sec:profiles_discussion}
    
        Regarding the temperature profiles shown in Figure~\ref{fig:phys_profiles} and presented in Section~\ref{sec:X-ray_interp}, we find a large discrepancy between our \textit{Chandra}+ALMA method and the results of \citet{Mantz2026}. At large radii, the temperature extrapolates to zero owing to the difference in slope between the gNFW and Vikhlinin profiles, with the electron pressure distribution steepening faster than the electron density. Even for gNFW-$\beta=3.5$, the temperature approaches zero at $\sim r_{200,c}$. We regard this feature as a direct consequence of the parametrisation rather than a physical effect. Future work performing a full joint-likelihood analysis combining X-ray and tSZ data could place priors on the temperature floor, properly accounting for this behaviour.
        
        The temperature distribution at small radii is also inconsistent with the \textit{XMM-Newton} measurements of \citet{Mantz2026} as well as the overall distribution of the temperature at intermediate redshifts \citep{McDonald2014}. The steepening of the electron temperature profile originates from the steep inner slope, $\gamma$, of the gNFW profile. Adopting the best-fitting A10 spherical model as the input for the X-ray analysis, we find a relatively cooler core, while still ``hot'' peaking at $12.5$\,keV. The hot core is consistent with the hot core visible in the projected temperature map (Figure~\ref{fig:maps}). Taken together, neither the standard Vikhlinin nor the isothermal temperature distribution adequately describes CL~J0459. A joint tSZ+X-ray analysis of a larger sample at these redshifts will be necessary to determine whether this holds for the broader population.
        
        The entropy distribution is more difficult to interpret. Classical modelling techniques, which combine a power-law density profile at large radii with an isothermal temperature, will by construction recover the canonical entropy profile: a rising power law in the outskirts with a flat core. With our method, the temperature and entropy are instead inferred from the electron pressure and density directly --- quantities more closely tied to the tSZ and X-ray surface brightness. Regardless of the different approach, we still find broad agreement with the population-averaged entropy profiles of \citet{McDonald2014}, showing a flat central entropy at the order of $200$~keV\,cm$^{2}$ rising radially, consistent with the self-similar profile $K \propto r^{1.1}$ \citep{Voit2005}. Constraining entropy from pressure and density rather than isothermal temperature fits will be crucial to test whether self-similar entropy profiles persist at high redshift. This is not expected as young clusters may not reach equilibrium given the ongoing accretion and mergers present at these epochs. Such dynamics could rather manifest as an isentropic but turbulent ICM.
             
    \subsection{Hydrostatic mass bias}\label{sec:mass_bias}

        \begin{figure}
            \centering
            \includegraphics[width=\linewidth]{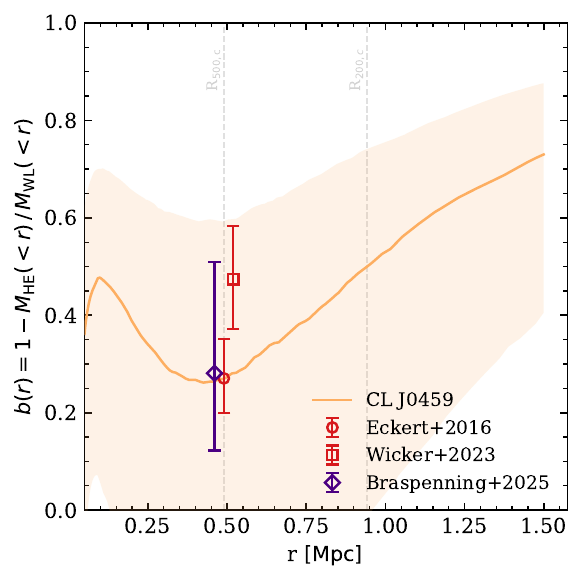}
            \caption{Hydrostatic mass bias as a function of radius with plotted 16$^{\rm th}-84^{\rm th}$ percentile range, computed using the $c$--$M$ relation of \citet{Diemer2019} and our hydrostatic mass estimate from resolved X-ray and tSZ measurements. The error on the hydrostatic mass bias is dominated by current weak lensing mass estimates from \citet{Zohren2022}. Results are compared to local estimates from \citet{Wicker2023} and \citet{Eckert2016} extrapolated to $z=1.7$, and to hydrodynamical simulations from \citet{Braspenning2025}.}
            \label{fig:HMbias}
        \end{figure}
    
        A promising diagnostic for an isentropic but turbulent ICM in an actively forming cluster is the hydrostatic mass bias \citep{Ghirardini2021, Braspenning2025}, with the hydrostatic mass defined as
        \begin{equation}\label{eq:HSE}
            M_{\rm HE}(<r) = -\frac{r^2}{\rho_{\rm gas}(<r) \, G} \frac{dP}{dr},
        \end{equation}
        with $\rho_{\rm gas}$ the enclosed gas density and $P$ the total pressure. The bias is then
        \begin{equation}
            b(r) = 1 - \frac{M_{\rm HE}(<r)}{M_{\rm tot}(<r)},
        \end{equation}
        where $M_{\rm tot}$ is estimated from weak lensing measurements. For CL~J0459, weak lensing masses were presented by \citet{Zohren2022}, who obtained a $2.5\sigma$ detection offset $\approx250$\,kpc from the X-ray centroid. Tangential reduced shear profiles were fit with spherical NFW models adopting the $c$--$M$ relation of \citet{Diemer2019}, yielding $M_{\rm 500,c,WL} = 3^{+5}_{-3} \times 10^{14}\,\rm M_\odot$.
        
        Using Equation~\eqref{eq:HSE}, we find $b_{\rm 500,c} = 1 - M_{\rm 500,c,HE}/M_{\rm 500,c,WL} = 0.27^{+0.32}_{-0.67}$. Figure~\ref{fig:HMbias} shows the radial bias profile, obtained by Monte Carlo sampling over the uncertainties, rejecting negative masses and treating the weak lensing uncertainties as a skewed Gaussian. Despite the large weak lensing uncertainties --- which could be substantially reduced with \textit{JWST} data \citep[e.g.,][]{Scofield2026} --- CL~J0459 is consistent with local bias estimates \citep{Eckert2016, Wicker2023} extrapolated to $z=1.7$ and with biases from hydrodynamical simulations at $\left<z\right>=1.5$ \citep{Braspenning2025}.
        
        The value of $b$ is on the high-side, in line with local clusters showing major mergers \citep[see, e.g.,][]{Umetsu2022}. The large bias, albeit with large error bars, is consistent with the interpretation of recent observations of protoclusters at slightly higher redshift. At $z \simeq 2$, \citet{Andreon2023}, \citet{DiMascolo2023}, and \citet{vanMarrewijk2024} found that dynamical masses from velocity dispersions are consistently $\approx3\times$ higher than tSZ-derived masses, suggesting the hot gas had insufficient time to provide full thermal pressure support which would lead to a large $b$. Whether this reflects cluster-to-cluster variation, a selection effect \citep[see discussion in][]{vanMarrewijk2024}, or a systematic evolution in thermal support can only be resolved with high-resolution X-ray and tSZ measurements of larger samples and weak lensing measurements at high $S/N$ out to $z\simeq2$.
            
        Furthermore, measurements of tSZ fluctuations might help determine the level of turbulence expected for young clusters \citep[see, e.g.,][]{Romero2025}. Given the high core entropy, its flat radial distribution, and the large hydrostatic mass bias, we expect this cluster to be turbulent. However, since high entropy does not necessarily lead to high velocity dispersion \citep[see, e.g.,][]{Eckert2025}, and with CL~J0459 likely experiencing an on-axis merger, bulk motions in the ICM can provide substantial non-thermal pressure support, reflected in the large hydrostatic mass bias (see discussion in \citealt{Umetsu2022}). With a larger sample --- and a better understanding of the systematics of measuring pressure fluctuations in interferometric data --- we can constrain the turbulence in these objects through tSZ measurements alone, providing an even more holistic approach than already presented in this work.
        
    \subsection{Scaling relations}\label{sec:scaling}
    
\begin{figure}
            \centering
            \includegraphics[width=0.94\linewidth]{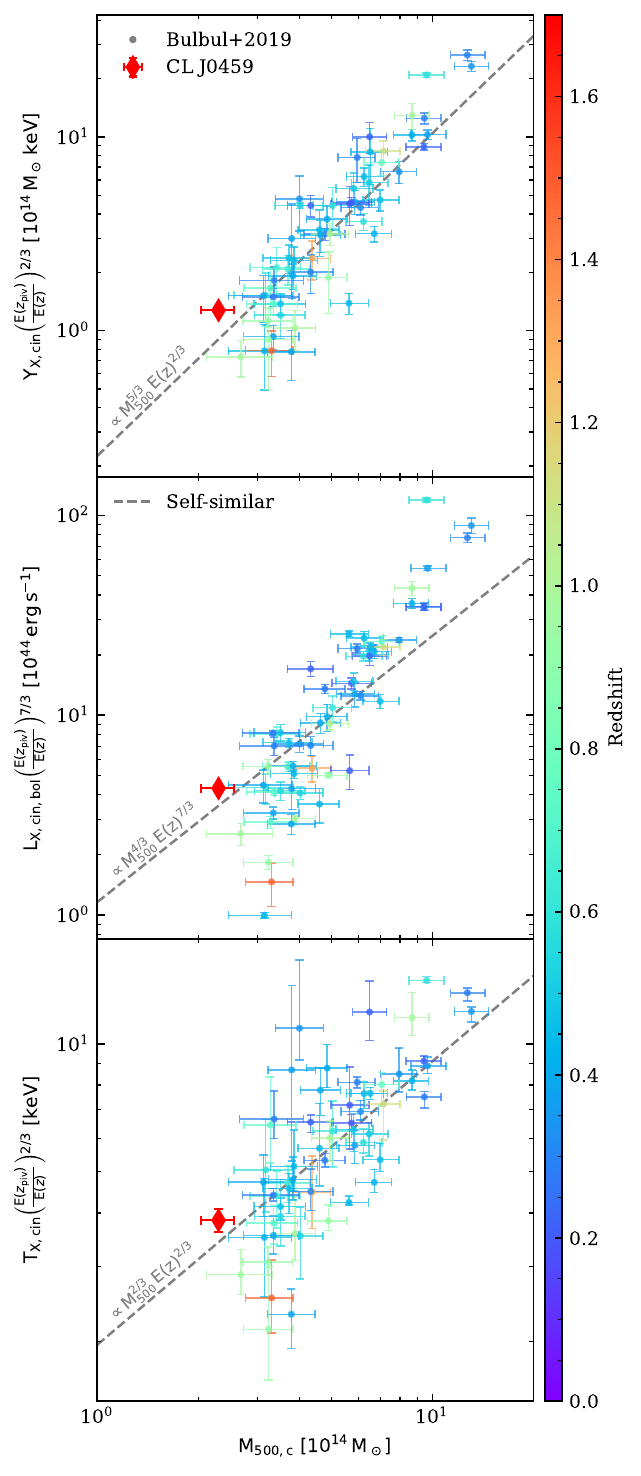}
            \caption{Scaling relations for CL J0459 (red diamond) compared to the SPT cluster sample of \citet{Bulbul2019}, colour-coded by redshift. Panels show E(z)-scaled core-included $Y_{X}$, bolometric $\rm L_X$, and spectroscopic temperature $\rm T_X$ versus $\rm M_{500,\mathrm{SZ}}$, all evaluated within $\rm R_{500,c}$ and scaled to pivot redshift $z_\mathrm{piv}=0.45$. Dashed lines show self-similar scaling.}
            \label{fig:scaling}
        \end{figure}

        As a final comparison, we place CL~J0459 in context with the sample of \citet{Bulbul2019}, an XMM-Newton programme targeting SPT-selected clusters at low to intermediate redshifts, shown together with CL~J0459 in Figure~\ref{fig:scaling}. Some simulations predict a break from self-similarity in the $Y$--$M$ scaling relation at high redshift, with lower $Y$ values at fixed mass compared to local samples \citep[e.g.,][]{Yu2015}. Whether this reflects differences in dynamical state and assembly history, or simply that $z \gtrsim 1.5$ clusters have systematically lower halo masses where a similar trend is observed \citep[e.g.,][]{Pop2022, Schaye2023, Aldas2026}, remains unclear. A tentative version of this behaviour with respect to redshift is visible in the \citet{Bulbul2019} sample, where higher-redshift clusters show consistently lower $Y$, $L_{\rm X}$, and $T_{\rm X}$ values relative to self-similar predictions. CL~J0459, however, falls slightly above the self-similar relations at the $1\sigma$-level, contrary to this trend. Here, we adopted the $\rm M_{500,c}$ derived from our A10-spherical model run as described in Section~\ref{sec:fit_ALMA} to be consistent with the halo mass definition in \citet{Bulbul2019} which are also tSZ-based. Although, we note that when adopting the $\rm M_{500,c}$ derived from SPT \citep{Bleem2024}, CL~J0459 would be put directly on the self-similar relations. Here the X-ray observables are derived from the \textit{Chandra}+ALMA fit of Section~\ref{sec:profiles}, meaning that $\rm Y_{\rm X} = Y$ for CL~J0459. 
        
        CL~J0459 appears self-similar in several respects --- its entropy profile, hydrostatic mass bias, and integrated observable scaling relations --- despite residing at $z=1.7$, an epoch when the matter-dominated cosmic expansion drives cluster growth and the short Hubble time implies ongoing accretion and correspondingly larger departures from equilibrium. his apparent regularity may, however, be misleading: CL~J0459 could be undergoing first core passage which can temporarily boost the $y$-value on time scales much smaller than the virialisation time \citep{wik2008}. If so, the self-similar appearance would reflect a transient, dynamic state rather than genuine dynamical relaxation, placing the cluster on or above the self-similar relations for reasons unrelated to virial equilibrium.
        
        Larger samples are needed to statistically characterise deviations from self-similarity and to probe the degree of thermalisation through hydrostatic mass bias measurements. This requires not only statistics, but high angular resolution ($\approx6\arcsec$) to resolve substructure in the tSZ and X-ray, accurate profile measurements, and sensitive optical data to constrain masses from probes that directly trace the total mass distribution.    

\section{Conclusions}\label{sec:conclusions}

    This paper demonstrates the power of high-resolution mm-wave data to derive thermodynamic distributions of young galaxy clusters by utilizing resolved measurements of the tSZ effect. Through a new methodology that jointly models the 2D surface brightness distributions of mm-wave and X-ray data, we measured radial electron pressure, density, temperature, and entropy profiles as well as projected thermodynamic maps of the ICM, using a combination of ACT, ALMA, \textit{Chandra}, and XMM-Newton data. We illustrated this on CL~J0459.6-4946 ($z=1.71$). Adding new, deep ALMA Band 1 and 3 measurements to the extensive archival data revealed the following:

    \begin{enumerate}
        \item ALMA Band 1 is uniquely well suited to study the ICM at ``X-ray-like'' angular resolutions in the mm-wave regime at high redshifts. With only a fraction of the observing time required in Band 3, ALMA Band 1 offers high-resolution mapping comparable to the main ALMA array in Band 3 with spatial scales as large as the one probed by Band 3 ACA but with $13\times$ its collecting area, a larger field of view, and lower system temperatures. The improved $uv$-coverage reduces our dependence on parametric profiles needed to interpolate and extrapolate the flux distribution, potentially paving the way towards less restrictive ICM parametrisations.
        \item By detecting CO line emission from cluster member galaxies with ALMA, we determined spectroscopic redshifts for our sources. We constrained the redshift of CL~J0459 to $z=1.71\pm0.02$, based on the median velocity distribution of 11 CO~$J$=2--1 emitters, two of which also have corresponding CO~$J$=1--0 detections.
        \item By coupling \texttt{MBProj2D} with our $uv$-based forward model, we modelled the 2D X-ray surface brightness distribution using the maximum likelihood radial pressure profile from the ALMA modelling. The ALMA+\textit{Chandra} analysis reproduced the thermodynamic profiles obtained from the $6\times$ deeper XMM-Newton data, even beyond $r_{500,c}$. The inferred profiles show a hot, high-entropy core declining radially, while the entropy follows a more self-similar behaviour out to $r_{500,c}$.
        \item The electron pressure and density profiles deviate from standard templates such as \citet{Arnaud2010} and \citet{McDonald2014}. The hot core and flat entropy distribution, combined with a relatively small effective path length $l_{\rm eff} \simeq 0.75\,r_{500,c}$ and a relatively large clumpiness factor of $C\simeq1.2$ within $r_{500,c}$, suggest that CL~J0459 has a compact, hot, dynamic core that is in the midst of thermalising its ICM. 
        This translates also to a relatively large hydrostatic mass bias albeit with large errorbars. HST weak lensing data \citep{Zohren2022} in combination with our improved hydrostatic mass estimate show that $b = 1 - M_{\rm 500,c, HE}/M_{\rm 500,c, WL} = 0.27^{+0.32}_{-0.67}$. Deeper JWST lensing observations will provide the S/N needed to test this case and constrain the hydrostatic mass bias as a function of radius.
        \item The ICM shows evidence for a cold, low-entropy, high-density region to the south of the cluster core. The absence of any corresponding perturbation in the Compton-$y$ map indicates this region is in pressure equilibrium, consistent with a sloshing cold front. Although its existence was previously posited \citep{McDonald2017}, our analysis, in combination with a large hydrostatic mass bias and the presence of an off-centred BCG at the location of the cold front, strengthens this interpretation. This is unprecedented at $z=1.7$, and demonstrates the unique capability of mm-wave instrumentation to push the study of cluster thermodynamics to high redshift.
    \end{enumerate}

    To summarize,  CL~J0459 has a disturbed ICM with a hot core that radially declines to lower temperatures. Although the temperature falls radially, the entropy distribution remains rather flat around $200$~keV\,cm$^{2}$ up to $r_{500,c}$. This nearly flat distribution with a clumpy ICM and a short effective path length suggest that CL~J0459 is dynamically active, possibly in the midst of thermalising its ICM by means of on-axis minor merger that is dissipating its kinetic energy to thermal energy. With any case study though it is hard to draw general conclusions, in our case about the overall thermal evolution of the cosmic web's most over dense regions. However, this paper provides a path forward: with current facilities, only a relatively short X-ray exposure (122~ks) is sufficient when combined with 3h-long ALMA measurements to fully characterise the ICM at high redshift. This opens a path towards studying how large-scale structure formation imprints on the thermodynamics of young and distant galaxy clusters for larger samples.

\section*{Acknowledgements}

    This paper makes use of the following ALMA data: ADS/JAO.ALMA\#2022.1.00427.S, ADS/JAO.ALMA\#2023.1.00213.S, and ADS/JAO.ALMA\#2023.1.01638.S ALMA is a partnership of ESO (representing its member states), NSF (USA) and NINS (Japan), together with NRC (Canada), MOST and ASIAA (Taiwan), and KASI (Republic of Korea), in cooperation with the Republic of Chile. The Joint ALMA Observatory is operated by ESO, AUI/NRAO and NAOJ.
    TM acknowledges support from the Agencia Estatal de Investigaci\'on (AEI) and the Ministerio de Ciencia, Innovaci\'on y Universidades (MICIU) Grant ATRAE2024-154740 funded by MICIU/AEI//10.13039/501100011033.
    TM is also partly supported by the Spanish program Unidad de Excelencia María de Maeztu CEX2020-001058-M, financed by MCIN/AEI//10.13039/501100011033, and by the MaX-CSIC Excellence Award MaX4-SOMMA-ICE.
    Support for ACT was through the U.S.~National Science Foundation through awards AST-0408698, AST-0965625, and AST-1440226 for the ACT project, as well as awards PHY-0355328, PHY-0855887 and PHY-1214379. Funding was also provided by Princeton University, the University of Pennsylvania, and a Canada Foundation for Innovation (CFI) award to UBC. ACT operated in the Parque Astron\'omico Atacama in northern Chile under the auspices of the Agencia Nacional de Investigaci\'on y Desarrollo (ANID). The development of multichroic detectors and lenses was supported by NASA grants NNX13AE56G and NNX14AB58G. Detector research at NIST was supported by the NIST Innovations in Measurement Science program. Computing for ACT was performed using the Princeton Research Computing resources at Princeton University, the National Energy Research Scientific Computing Center (NERSC), and the Niagara supercomputer at the SciNet HPC Consortium. SciNet is funded by the CFI under the auspices of Compute Canada, the Government of Ontario, the Ontario Research Fund–Research Excellence, and the University of Toronto. We thank the Republic of Chile for hosting ACT in the northern Atacama, and the local indigenous Licanantay communities whom we follow in observing and learning from the night sky.
    JvM acknowledges assistance from Allegro, the European ALMA Regional Center node in the Netherlands.
    This work was supported by JSPS KAKENHI Grant Numbers JP25K07372 (TK), JP25K23398 (SU), and JP26K00741 (SU).
    SU acknowledges support by Program for Forming Japan’s Peak Research Universities (J-PEAKS) Grant Number JPJS00420230006.
    FAG acknowledges support from the ANID BASAL project FB210003, from the ANID FONDECYT Regular grant 1251493 and from the HORIZON-MSCA-2021-SE-01 Research and Innovation Programme under the Marie Sklodowska Curie grant agreement number 101086388
    MH acknowledges support from the National Research Foundation of South Africa (grant nos. CPRR240513218388, SKRG20200101-097792).
    KM acknowledges support from the National Research Foundation of South Africa.
    During the preparation of this work, JvM used Claude (Anthropic) to assist with code development for the analysis, draft figure captions and spell and grammar check the text. The authors reviewed and edited all output and take full responsibility for the content of the publication.

\section*{Data Availability}

    The ALMA data are available from the ALMA Science Archive (projects
    \texttt{2022.1.00427.S}, \texttt{2023.1.00213.S} and
    \texttt{2023.1.01638.S}), the \textit{Chandra} data from the Chandra Data
    Archive (ObsIDs 17211, 17501, 17502, 18711, 18824 and 18853), the
    XMM-Newton data from the XMM-Newton Science Archive (ObsID 0801950101),
    and the ACT maps from LAMBDA\footnote{\url{https://lambda.gsfc.nasa.gov}}.
    Derived products, including posterior samples and thermodynamic profiles,
    are available on reasonable request to the corresponding author. The
    analysis code will be made public in due time.

\bibliographystyle{mnras}
\bibliography{J0459}

\affiliations


\appendix

    \section{Anomaly in the Band 1 data}\label{app:LO}

        \begin{figure*}
            \centering
            \includegraphics[width=\linewidth]{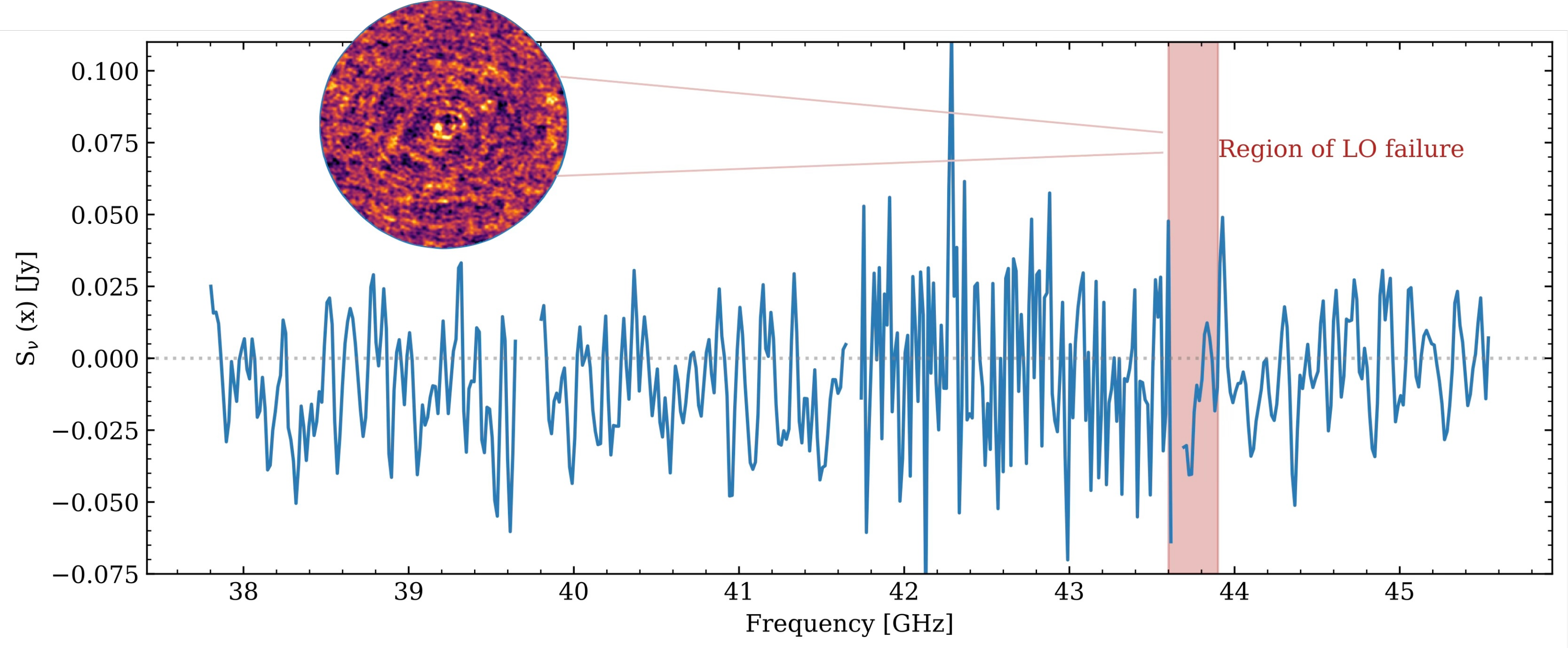}
            \caption{Spectral line profile (solid) of the Band 1 data at the location of the brightest line detected in Band 3. We further highlight the spectral region where the LO spikes, showing the channel map at a location of $\nu= 43.9140~$GHz. The profile is made using 15 MHz ($\approx 110$ km/s) binning adopting a central frequency corresponding to the expected location of the CO J=1-0 line for a redshift of $z=1.71$.}
            \label{fig:bad_slp}
        \end{figure*}

        Here, we highlight a region in Figure~\ref{fig:bad_slp}, in which we identified a feature that is likely related to a local oscillator (LO) leakage problem \citep[see the ALMA technical Handbook][]{Remijan2019}, a known issue for Band 1 data. The affected channels are therefore removed from the visibilities before further analysis. After this feature was flagged, ALMA now prohibits observations taken place at this exact frequency.
    
    \section{Cluster Members}\label{app:members}
    
        \begin{figure}
            \centering
            \includegraphics[width=\linewidth]{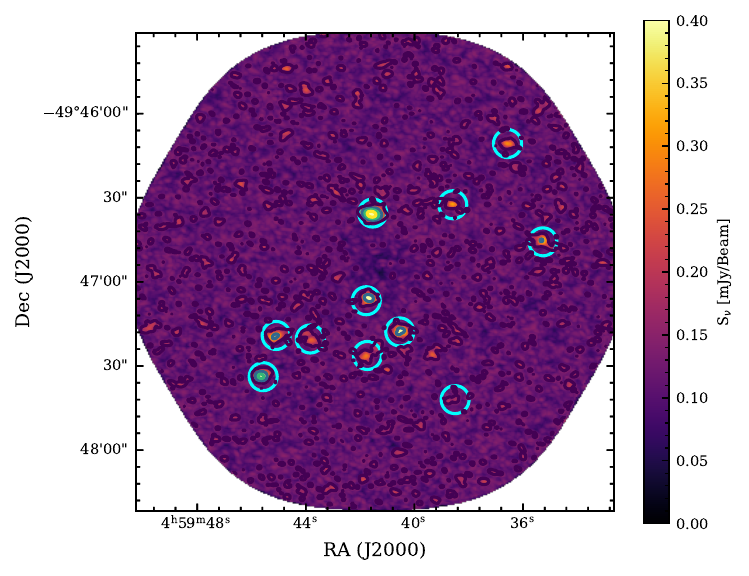}
            \caption{Moment-8 map of the jointly imaged ACA and 12m-array Band 3 observations of CL~J0459. Contours are drawn at 0.15, 0.35, 0.55, 0.75 mJy/beam. Circles are masked regions of line emitting sources, used to extract the spectrum. }
            \label{fig:moment8_}
        \end{figure}

        \begin{table}
            \centering
            \caption{Mm-bright continuum source fluxes and spectral indices.}
            \label{tab:cont_fluxes}
            \setlength{\tabcolsep}{3pt}
            \begin{tabular}{rrrrr}
                \toprule
                RA           & Dec         & $S_{\rm B3}^{\,a}$   & $S_{\rm B1}^{\,a,b}$ & $\alpha_\nu^{\,c}$   \\
                {[$^\circ$]} & [$^\circ$]  & [$\mu$Jy]            & [$\mu$Jy]            &                      \\
                \midrule
                $74.91807$   & $-49.79037$ & $38.0^{+5.3}_{-5.4}$ & $11.4^{+3.4}_{-3.2}$ & $1.5^{+0.5}_{-0.4}$  \\
                $74.94705$   & $-49.77050$ & $110^{+19}_{-18}$    & $184^{+12}_{-11}$    & $-0.7^{+0.2}_{-0.2}$ \\
                $74.93206$   & $-49.78899$ & $12.7^{+5.2}_{-5.6}$ & $<11.3$              & --                   \\
                $74.89715$   & $-49.77915$ & $38^{+11}_{-11}$     & $<21.0^d$            & --                   \\
                $74.92427$   & $-49.77154$ & $16.8^{+5.7}_{-5.4}$ & $26.7^{+5.1}_{-5.0}$ & $-0.6^{+0.4}_{-0.5}$ \\
                \bottomrule
            \end{tabular}
            \begin{minipage}{\columnwidth}
                \textbf{Notes.} $^a$ Intrinsic model fluxes at 92 and 41.7\,GHz: posterior median and 68 per cent interval of the forward-modelled point-source fit. $^b$ Upper limits are 95 per cent for S/N $<$ 3. $^c$ Two-point spectral index between 41.7 and 92\,GHz. $^d$ Located at half the primary beam of Band 1, which roughly doubles the intrinsic noise level.
            \end{minipage}
        \end{table}

        \begin{table*}
            \centering
            \caption{Line properties of compact sources towards CL~J0459.}
            \begin{tabular}{cccccc}
                \toprule
                RA [$^\circ$] & Dec [$^\circ$] & $\nu_{\rm obs}$ [GHz] & $z$ & FWHM [km\,s$^{-1}$] & $S\Delta v$ [mJy\,km\,s$^{-1}$] \\
                \midrule
                \multicolumn{6}{l}{\textbf{ALMA Band 3}} \\
                \midrule
                $74.91078$ & $-49.77570$ & $84.121 \pm 0.004$ & $1.74055 \pm 0.00014$ & $160 \pm 40$ & $53 \pm 11$ \\
                $74.93789$ & $-49.78865$ & $84.1619 \pm 0.0030$ & $1.73922 \pm 0.00010$ & $241 \pm 25$ & $134 \pm 12$ \\
                $74.93265$ & $-49.78900$ & $84.399 \pm 0.008$ & $1.73151 \pm 0.00026$ & $500 \pm 70$ & $141 \pm 16$ \\
                $74.91042$ & $-49.79501$ & $84.496 \pm 0.006$ & $1.72840 \pm 0.00019$ & $180 \pm 50$ & $49 \pm 11$ \\
                $74.92307$ & $-49.77652$ & $84.5658 \pm 0.0015$ & $1.72614 \pm 0.00005$ & $205 \pm 12$ & $305 \pm 16$ \\
                $74.91891$ & $-49.78830$ & $84.853 \pm 0.005$ & $1.71692 \pm 0.00015$ & $560 \pm 40$ & $311 \pm 19$ \\
                $74.93988$ & $-49.79273$ & $85.1739 \pm 0.0023$ & $1.70667 \pm 0.00007$ & $235 \pm 19$ & $170 \pm 12$ \\
                $74.92406$ & $-49.78521$ & $85.369 \pm 0.006$ & $1.70049 \pm 0.00019$ & $480 \pm 50$ & $246 \pm 22$ \\
                $74.92388$ & $-49.79063$ & $85.351 \pm 0.006$ & $1.70105 \pm 0.00019$ & $210 \pm 50$ & $63 \pm 13$ \\
                $74.89696$ & $-49.77937$ & $85.834 \pm 0.009$ & $1.68585 \pm 0.00029$ & $550 \pm 80$ & $167 \pm 20$ \\
                $74.90239$ & $-49.76963$ & $85.826 \pm 0.007$ & $1.68610 \pm 0.00023$ & $480 \pm 60$ & $149 \pm 17$ \\
                \midrule
                \multicolumn{6}{l}{\textbf{ALMA Band 1}} \\
                \midrule
                $74.92307$ & $-49.77652$ & $42.2831 \pm 0.0030$ & $1.72618 \pm 0.00019$ & $170 \pm 50$ & $65 \pm 15$ \\
                $74.93988$ & $-49.79273$ & $42.5848 \pm 0.0020$ & $1.70686 \pm 0.00013$ & $170 \pm 24$ & $70 \pm 10$ \\
                $74.91843$ & $-49.79029$ & $41.499 \pm 0.008$ & -- & $580 \pm 130$ & $117 \pm 23$ \\
                $74.94688$ & $-49.77029$ & $41.250 \pm 0.017$ & -- & $670 \pm 290$ & $73 \pm 27$ \\
                \bottomrule
            \end{tabular}
            \label{tab:lines}
            \begin{minipage}{\textwidth}
                \textbf{Notes.} Errors are $1\sigma$ uncertainties from the fit covariance, adopting a single Gaussian per line profile, rounded following the Particle Data Group convention. Redshifts assume CO(2--1) in Band 3 and CO(1--0) in Band 1. Band 1 sources 1 and 2 are the same galaxies as Band 3 sources 5 and 7, respectively, with velocity offsets (Band 1 $-$ Band 3) of $4 \pm 22$ and $21 \pm 16$~km\,s$^{-1}$. The remaining two Band 1 line emitters have no Band 3 counterpart; their redshifts are omitted since they are most likely interlopers.
            \end{minipage}
        \end{table*}

        Here, we provide additional details on the cluster members identified in the Band~1~\&~~3 data for legacy purposes (see Section~\ref{sec:line_sources}). Table~\ref{tab:cont_fluxes} lists the continuum flux properties of the compact sources and Table~\ref{tab:lines} lists the fitted line parameters both with their $1\sigma$ uncertainties. The line fluxes are estimated by modelling each spectral line with a Gaussian profile. Figure~\ref{fig:moment8_} shows the moment--8 map of the Band~3 data, a two--dimensional projection in which each pixel value corresponds to the maximum intensity along the frequency axis. This map highlights the locations of the galaxies together with the masks used to extract the individual spectral profiles. 

    \section{Analytical derivation of \texorpdfstring{$\ell_{\mathrm{eff}}$}{leff}}\label{app:leff}
    
        Following the late 90's and early 2000's work on deriving $H_0$ via a joint X-ray-SZ analysis \citep[see, e.g.,][]{Birkinshaw1991, Hughes1998, Reese2002}, we decided to use an analytical expression of $l_{\rm eff}$ to validate the accuracy of adopting a constant value of $l_{\rm eff}$ and investigate the systematic uncertainties associated with the adopted parametrisation. The full derivation of $l_{\rm eff}$ when assuming a $\beta-$profile best-describes the ICM density distribution, $n_{\rm e}$, is given below;
        
        The $\beta$-profile describes the electron density as a function of three-dimensional radius $r$ as
        \begin{equation}\label{eq:beta_profile_3d}
            n_\text{e}(r) = n_0\left[1+\left(\frac{r}{r_\text{c}}\right)^2\right]^{-3\beta/2},
        \end{equation}
        where $n_0$ is the central electron density, $r_\text{c}$ is the core radius, and $\beta$ describes the slope of the profile. Adopting this profile, the projected ratio $\int n_\text{e}^2\,\text{d}z / \int n_\text{e}\,\text{d}z$ has the analytical form
        \begin{equation}\label{eq:beta_profile}
            \frac{\int n_\text{e}^2\,\text{d}z}{\int n_\text{e}\,\text{d}z} = n_0\left[1+\left(\frac{R}{r_\text{c}}\right)^2\right]^{-3\beta/2},
        \end{equation}
        in Cartesian coordinates with $r^2 = x^2 + y^2 + z^2$ and $R^2 = x^2 + y^2$. To evaluate Eq.~\eqref{eq:leff} analytically, we make use of the standard definite integral
        \begin{equation}\label{eq:standard_integral}
            \int_{0}^{+\infty} \left(A^2+u^2\right)^{-pm}\text{d}u = \frac{\sqrt{\pi}}{2}\frac{\Gamma(pm-1/2)}{\Gamma(pm)}A^{1-2pm},
        \end{equation}
        where $\Gamma(\cdot)$ denotes the Gamma function. Substituting $A^2 = 1 + R^2/r_\text{c}^2$, $u = z/r_\text{c}$, $m = 3\beta/2$, and setting $p = 1$ in the numerator and $p = 2$ in the denominator of Eq.~\eqref{eq:leff}, we obtain
        \begin{align*}
            l_\text{eff}
            &= \frac{\left[\int_{0}^{+\infty} n_\text{e}\,\mathrm{d}z\right]^2}{\int_{0}^{+\infty} n_\text{e}^2\,\mathrm{d}z} \\[1ex]
            &= \frac{\left[\int_0^{\infty}\left(1+\frac{R^2}{r_\text{c}^2}+\frac{z^2}{r_\text{c}^2}\right)^{-3\beta/2}\mathrm{d}z\right]^2}{\int_0^{\infty}\left(1+\frac{R^2}{r_\text{c}^2}+\frac{z^2}{r_\text{c}^2}\right)^{-3\beta}\mathrm{d}z} \\[1ex]
            &= \frac{\left[\int_0^{+\infty}\left(A^2+u^2\right)^{-m}\mathrm{d}u\right]^2}{\int_0^{+\infty}\left(A^2+u^2\right)^{-2m}\mathrm{d}u}.
        \end{align*}
        Applying Eq.~\eqref{eq:standard_integral} to both integrals and simplifying yields
        \begin{align*}
            l_\text{eff}
            &= r_\text{c}\frac{\left[\frac{\sqrt{\pi}}{2}\frac{\Gamma(m-1/2)}{\Gamma(m)}A^{1-2m}\right]^2}{\frac{\sqrt{\pi}}{2}\frac{\Gamma(2m-1/2)}{\Gamma(2m)}A^{1-4m}} \\[1ex]
            &= \frac{r_\text{c}\sqrt{\pi}}{2}A\frac{\Gamma\left(m-1/2\right)^2\Gamma(2m)}{\Gamma(m)^2\Gamma(2m-1/2)}.
        \end{align*}
        Substituting back $A = \left(1 + R^2/r_\text{c}^2\right)^{1/2}$ and $m = 3\beta/2$ gives the final result,
        \begin{equation}\label{eq:l_eff_projected_app}
            l_\text{eff} = \frac{r_\text{c}\sqrt{\pi}}{2}\left(1+\frac{R^2}{r_\text{c}^2}\right)^{1/2}\frac{\Gamma\left(3\beta/2-1/2\right)^2\Gamma(3\beta)}{\Gamma(3\beta/2)^2\Gamma(3\beta-1/2)},
        \end{equation}
        with $\beta > 1/3$. Interestingly, the spatial dependence of $l_{\rm eff}$ is governed entirely by the factor $\left(1+R^2/r_\text{c}^2\right)^{1/2}$, with $\beta$ entering only as amplitude parameters.

\bsp	
\label{lastpage}
\end{document}